\documentclass[12pt]{article}
\usepackage{latexsym}
\usepackage{graphics}
\usepackage{epsfig,amssymb,amsmath,graphicx,amsfonts}
\usepackage[dvips]{color}

\makeatletter
\newcommand{\singlespacing}{\let\CS=\@currsize\renewcommand{\baselinestretch}{1}\tiny\CS}
\newcommand{\oneandahalfspacing}{\let\CS=\@currsize\renewcommand{\baselinestretch}{1.25}\tiny\CS}
\newcommand{\doublespacing}{\let\CS=\@currsize\renewcommand{\baselinestretch}{1.5}\tiny\CS}

\newcommand{\bc}{\begin{center}}
\newcommand{\ec}{\end{center}}
\newcommand{\bfl}{\begin{flushleft}}
\newcommand{\efl}{\end{flushleft}}

\newcommand{\beqa}{\begin{eqnarray}}
\newcommand{\eeqa}{\end{eqnarray}}
\newcommand{\beqan}{\begin{eqnarray*}}
\newcommand{\eeqan}{\end{eqnarray*}}
\newcommand{\beq}{\begin{equation}}
\newcommand{\eeq}{\end{equation}}
\newcommand{\beqn}{\begin{equation*}}
\newcommand{\eeqn}{\end{equation*}}

\newcommand{\real}{{\mathcal {R}}}

\newcommand{\expect}{{\Bbb {E}}}

\newtheorem{theorem}            {Theorem}
\newtheorem{corollary}          [theorem]{Corollary}

\newtheorem{definition}         [theorem]{Definition}

\newtheorem{lemma}              [theorem]{Lemma}

\newcommand{\bM}{\mathbf{M}}
\newcommand{\bX}{\mathbf{X}}
\newcommand{\bY}{\mathbf{Y}}
\newcommand{\bI}{\mathbf{I}}
\newcommand{\bJ}{\mathbf{J}}
\newcommand{\bZ}{\mathbf{Z}}
\newcommand{\bK}{\mathbf{K}}
\newcommand{\bS}{\mathbf{S}}
\newcommand{\bA}{\mathbf{A}}
\newcommand{\bB}{\mathbf{B}}
\newcommand{\bC}{\mathbf{C}}
\newcommand{\bD}{\mathbf{D}}
\newcommand{\bQ}{\mathbf{Q}}
\newcommand{\bU}{\mathbf{U}}
\newcommand{\bV}{\mathbf{V}}
\newcommand{\bv}{\mathbf{v}}
\newcommand{\bW}{\mathbf{W}}

\newcommand{\Cov}{\mathrm{Cov}}
\newcommand{\Tr}{\mathrm{Tr}}
\newcommand{\Diag}{\mathrm{Diag}}
\newcommand{\Var}{\mathrm{Var}}

\begin{document}
\title{An Extremal Inequality Motivated by Multiterminal Information Theoretic Problems}
\author{Tie Liu and Pramod Viswanath\thanks{Tie Liu is with the Department of Electrical
and Computer Engineering at the Texas A\&M University, College
Station, TX 77843, USA (e-mail: {\tt tieliu@ece.tamu.edu}). Pramod
Viswanath is with the Department of Electrical and Computer
Engineering at the University of Illinois at Urbana-Champaign,
Urbana, IL 61801, USA (e-mail: {\tt pramodv@uiuc.edu}).}}

\date{\today}

\maketitle

\begin{abstract}
We prove a new extremal inequality, motivated by the vector Gaussian
broadcast channel and the distributed source coding with a single
quadratic distortion constraint problems. As a corollary, this
inequality yields a generalization of the classical entropy-power
inequality (EPI). As another corollary, this inequality sheds
insight into maximizing the differential entropy of the sum of two
dependent random variables.
\end{abstract}


\section{Introduction} \label{sec:intro}
Like many other important results in information theory, the
classical entropy-power inequality (EPI) was discovered by Shannon
\cite{Shannon48} (even though the first rigorous proof was given by
Stam \cite{Stam59} and was later simplified by Blachman
\cite{Blachman65}). In \cite[p.~641]{Shannon48}, Shannon used the
EPI to prove a lower bound on the capacity of additive noise
channels. While this first application was on a point-to-point
scenario, the real value of the EPI showed up much later in the
multiterminal source/channel coding problems where the tension among
users of different interests cannot be resolved by Fano's inequality
alone. The most celebrated examples include Bergman's solution
\cite{Bergman74} to the scalar Gaussian broadcast channel problem,
Oohama's solution \cite{Oohama98} to the scalar quadratic Gaussian
CEO problem, and Ozarow's solution \cite{Ozarow80} to the scalar
Gaussian two-description problem.

Denote the set of real numbers by $\real$. Let $\bX$, $\bZ$ be two
independent random vectors with densities in $\real^n$. The
classical EPI states that \beq
\exp\left[\frac{2}{n}h(\bX+\bZ)\right] \geq
\exp\left[\frac{2}{n}h(\bX)\right] +
\exp\left[\frac{2}{n}h(\bZ)\right]. \label{eq:EPI} \eeq Here
$h(\bX)$ denotes the differential entropy of $\bX$, and the equality
holds if and only if $\bX$, $\bZ$ are Gaussian \emph{and with
proportional covariance matrices}.

Fix $\bZ$ to be Gaussian with covariance matrix $\bK_Z$. Assume that
$\bK_Z$ is strictly positive definite. Consider the optimization
problem \beq \max_{p(\mathbf{x})} \,
\left\{h(\bX)-\mu{h}(\bX+\bZ)\right\}, \label{eq:Opt-UNC} \eeq where
$\mu \in \real$, and the maximization is over all random vector
$\bX$ independent of $\bZ$. The classical EPI can be used to show
that for any $\mu > 1$, a Gaussian $\bX$ with a covariance matrix
proportional to $\bK_Z$ is an optimal solution of this optimization
problem. This can be done as follows. By the classical EPI, \beq
h(\bX)-\mu h(\bX+\bZ) \leq h(\bX)-\frac{\mu
n}{2}\log\left(\exp\left[\frac{2}{n}h(\bX)\right]+\exp\left[\frac{2}{n}h(\bZ)\right]\right).
\label{eq:a} \eeq For any fixed $a \in \real$ and $\mu > 1$, the
function \beq f(t;a) = t-\frac{\mu
n}{2}\log\left(\exp\left[\frac{2}{n}t\right]+\exp\left[\frac{2}{n}a\right]\right),
\label{eq:f} \eeq is concave in $t$ and has a global maxima at \beq
t = a-\frac{n}{2}\log(\mu-1). \label{eq:t} \eeq Hence the right-hand
side of \eqref{eq:a} can be further bounded from above as \beq
h(\bX)-\frac{\mu
n}{2}\log\left(\exp\left[\frac{2}{n}h(\bX)\right]+\exp\left[\frac{2}{n}h(\bZ)\right]\right)
\leq f\left(h(\bZ)-\frac{n}{2}\log(\mu-1);h(\bZ)\right).
\label{eq:b} \eeq The equality conditions of \eqref{eq:a} and
\eqref{eq:b} imply that a Gaussian $\bX$ with covariance matrix
$(\mu-1)^{-1}\bK_Z$ is an optimal solution of the optimization
problem \eqref{eq:Opt-UNC}.

Note that in solving the above optimization problem, the classical
EPI not only forces the optimal solution to be Gaussian, but also
imposes a certain covariance structure on the Gaussian optimal
solution. Hence a natural question to ask is what happens if there
is an extra covariance constraint such that the original Gaussian
optimal solution is no longer admissible. In that case, the
classical EPI can still be used; however, the equality condition may
no longer be met by the new optimal Gaussian solution because it may
no longer have the required proportionality. In particular, one
would be interested in finding out whether under the extra
covariance constraint, a \emph{Gaussian} $\bX$ is still an optimal
solution to optimization problems such as \eqref{eq:Opt-UNC}.

One particular type of covariance constraint is the following matrix
covariance constraint: \beq \Cov(\bX) \preceq \bS. \eeq Here
$\Cov(\bX)$ denotes the covariance matrix of $\bX$, ``$\preceq$"
represents ``less or equal to" in the positive semidefinite partial
ordering of real symmetric matrices, and $\bS$ is a positive
semidefinite matrix. The reason for considering such a matrix
covariance constraint is largely due to its generality: it subsumes
many other covariance constraints including the important trace
constraint.

The focus of this paper is the following slightly more general
optimization problem:
\beq \begin{array}{ll} \max_{p(\mathbf{x})} & h(\bX+\bZ_1)-\mu h(\bX+\bZ_2) \\
\mathrm{subject\;to} & \Cov(\bX) \preceq \bS, \end{array}
\label{eq:Opt-G} \eeq where $\bZ_1$, $\bZ_2$ are Gaussian vectors
with strictly positive definite covariance matrix $\bK_{Z_1}$ and
$\bK_{Z_2}$, respectively, and the maximization is over all random
vector $\bX$ independent of $\bZ_1$ and $\bZ_2$. As we shall see,
such an optimization problem appears naturally when one is to
evaluate certain genie-aided outer bounds on the capacity/rate
region for the vector Gaussian broadcast channel and the distributed
source coding with a single quadratic distortion constraint
problems. Our main result is summarized in the following theorem.

\begin{theorem} \label{theorem:main}
For any $\mu \geq 1$ and any positive semidefinite $\bS$, a Gaussian
$\bX$ is an optimal solution of the optimization problem
\eqref{eq:Opt-G}.
\end{theorem}

The rest of the paper is organized as follows. In
Section~\ref{sec:main}, we prove our main result. We give two
proofs: a direct proof using the classical EPI, and a strengthened
proof following the perturbation approach of Stam \cite{Stam59} and
Blachman \cite{Blachman65}. In Section~\ref{sec:ram}, we discuss
some ramifications of the main result. In Section~\ref{sec:motiv},
we apply our main result to the vector Gaussian broadcast channel
and the distributed source coding with a single quadratic distortion
constraint problems. For the former problem, our main result leads
to an \emph{exact} characterization of the capacity region. Finally,
in Section~\ref{sec:con}, we conclude by summarizing our
contribution in the context of the applications of information
theoretic inequalities in resolving multiterminal
transmission/compression problems.

\section{Proofs of the Main Result} \label{sec:main}
\subsection{A Direct Proof}
In this first proof, we show that the classical EPI can be
\emph{appropriately} used to give a direct proof to
Theorem~\ref{theorem:main}. The fact that the classical EPI is
relevant here is not surprising, considering that the objective
function of the optimization problem \eqref{eq:Opt-G} involves the
entropy of the sum of two independent random vectors. Nonetheless,
based on our discussion in Section~\ref{sec:intro}, a direct use of
the classical EPI might be loose because the covariance matrix of
the optimal Gaussian solution might not have the required
proportionality.

Our approach to resolve this issue is inspired by the mathematical
import of an interesting technique, called \emph{enhancement},
introduced by Weingarten et al. \cite{WSS05}. Our proof combines the
idea of enhancement with the worst additive noise lemma
\cite{Ihara78}, \cite[Lemma II.2]{DC01} stated as follows.

\begin{lemma}[Worst additive noise lemma]
\label{lemma:worst} Let $\bZ$ be a Gaussian vector with covariance
matrix $\bK_Z$, and let $\bK_X$ be a positive semidefinite matrix.
Consider the following optimization problem:
\beq \begin{array}{ll} \min_{p(\mathbf{x})} & I(\bZ;\bZ+\bX) \\
\mathrm{subject\;to} & \Cov(\bX) = \bK_X, \end{array}
\label{eq:Opt-W} \eeq where $I(\bZ;\bZ+\bX)$ denotes the mutual
information between $\bZ$ and $\bX+\bZ$, and the maximization is
over all random vector $\bX$ independent of $\bZ$. A Gaussian $\bX$
is an optimal solution of this optimization problem (no matter
$\bK_X$ and $\bK_Z$ are proportional or not).
\end{lemma}

The details of the direct proof are in Appendix
\ref{app:basicproof}.

\subsection{A Perturbation Proof}
From the optimization theoretic point of view, the power of the
classical EPI lies in its ability to find global optima in nonconvex
optimization problems such as (\ref{eq:Opt-UNC}). Hence one can
imagine that proof of the classical EPI cannot be accomplished by
any local optimization techniques. Indeed, in their classical proof
Stam \cite{Stam59} and Blachman \cite{Blachman65} used a
\emph{perturbation} approach, which amounts to find a
\emph{monotone} path from any distributions of the participating
random vectors (i.e., $\bX$ and $\bZ$ in \eqref{eq:EPI}) to the
optimal distributions (Gaussian distributions with proportional
covariance matrices) for which the classical EPI holds with
equality. The monotonicity guarantees that \emph{any} distributions
along the path satisfy the desired inequality, and hence the ones to
begin with. A different perturbation was later used by Dembo et al.
\cite[p.~1509]{DCT91}. The main idea, however, remains the same as
that of Stam and Blachman's.

Proving monotonicity needs isoperimetric inequalities. In case of
the classical EPI, it needs the classical Fisher information
inequality (FII) \cite[Theorem~13]{DCT91}. Fisher information is an
important quantity in statistical estimation theory. An interesting
estimation theoretic proof using the data processing inequality for
Fisher information was given by Zamir \cite{Zamir98}. (The classical
FII can also be proved by using the standard data processing
inequality for mutual information, invoking a connection between
Fisher information and mutual information explicitly established by
Guo et al. \cite[Corollary~2]{GSV05}.) This connection between the
EPI and the FII is usually thought of as the estimation view of the
classical EPI.

We can use the perturbation idea to give a stronger proof to
Theorem~\ref{theorem:main}. We construct a monotone path using the
``covariance-preserving" transformation, which was previously used
by Dembo et al. \cite[p.~1509]{DCT91} in their perturbation proof of
the classical EPI. To prove the monotonicity, we need the following
results on Fisher information matrix.

\begin{lemma} \label{lemma:FI} Denote by $\bJ(\bX)$ the Fisher
information matrix of random vector $\bX$.
\begin{enumerate}
\item (Cram\'{e}r-Rao inequality) For any random vector $\bU$
(of which the Fisher information matrix is well defined) with a
strictly positive definite covariance matrix, \beq \bJ(\bU) \succeq
\Cov^{-1}(\bU). \eeq
\item (Matrix FII) For any independent random vectors $\bU$, $\bV$
and any square matrix $\bA$, \beq \bJ(\bU+\bV) \preceq
\bA\bJ(\bU)\bA^t+(\bI-\bA)\bJ(\bV)(\bI-\bA)^t. \eeq Here $\bI$ is
the identity matrix.
\end{enumerate}
\end{lemma}

For completeness, a proof of the above lemma using the properties of
\emph{score function} is provided in Appendix \ref{app:myFII}. The
details of the perturbation proof are in Appendix
\ref{app:perturbproof}.

\section{Ramifications of the Main Result} \label{sec:ram}
In this section, we discuss two special cases of the optimization
problem \eqref{eq:Opt-G} to demonstrate the breadth of our main
result. We term these two scenarios as the \emph{degraded} case and
the \emph{extremely-skewed} case. By considering the degraded case,
we prove a generalization of the classical EPI. By considering the
extremely-skewed case, we establish a connection between our result
and the classical result of Cover and Zhang \cite{CZ94} on the
maximum differential entropy of the sum of two \emph{dependent}
random variables.

\subsection{The Degraded Case}
In the degraded case, we have either $\bK_{Z_1} \preceq \bK_{Z_2}$
or $\bK_{Z_1} \succeq \bK_{Z_2}$. First consider the case $\bK_{Z_1}
\preceq \bK_{Z_2}$. We have the following results.

\begin{corollary} \label{cor:dcr}
Let $\bZ_1$, $\bZ$ be two independent Gaussian vectors with
covariance matrix $\bK_{Z_1}$ and $\bK_Z$, respectively. Assume that
$\bK_{Z_1}$ is strictly positive definite. Consider the following
optimization problem:
\beq \begin{array}{ll} \max_{p(\mathbf{x})} & h(\bX+\bZ_1)-\mu h(\bX+\bZ_1+\bZ) \\
\mathrm{subject\;to} & \Cov(\bX) \preceq \bS, \end{array}
\label{opt:dcr} \eeq where the maximization is over all random
vector $\bX$ independent of $\bZ_1$ and $\bZ$. For any $\mu \in
\real$ and any positive semidefinite $\bS$, a Gaussian $\bX$ is an
optimal solution of this optimization problem.
\end{corollary}

\emph{Proof.} For $\mu \geq 1$, the corollary is a special case of
Theorem \ref{theorem:main} with $\bZ_2=\bZ_1+\bZ$. For $\mu \leq 0$,
the corollary also holds because $h(\bX+\bZ_1)$ and
$h(\bX+\bZ_1+\bZ)$ are \emph{simultaneously} maximized when $\bX$ is
Gaussian with covariance matrix $\bS$. This left us the only case
where $\mu \in (0,1)$, which we prove next.

The objective function of optimization problem \eqref{opt:dcr} can
be written as \beq (1-\mu)h(\bX+\bZ_1)-\mu I(\bZ;\bX+\bZ_1+\bZ).
\label{eq:dcr} \eeq Here $h(\bX+\bZ_1)$ is maximized when $\bX$ is
Gaussian with covariance matrix $\bS$. By the worst noise result of
Lemma \ref{lemma:worst}, $I(\bZ;\bX+\bZ_1+\bZ)$ is minimized when
$\bX$ is Gaussian. Further within the Gaussian class, the one with
the full covariance matrix $\bS$ minimizes $I(\bZ;\bZ+\bX+\bZ_1)$.
For $\mu \in (0,1)$, both $\mu$ and $1-\mu$ are positive. We
conclude that the objective function (\ref{eq:dcr}) is maximized
when $\bX$ is Gaussian with covariance matrix $\bS$. This completes
the proof. \hfill $\square$

\begin{corollary}\label{cor:dcr2}
Let $\bZ$ be a Gaussian vector with covariance matrix $\bK_Z$.
Assume that $\bK_Z$ is strictly positive definite. Consider the
following optimization problem
\beq \begin{array}{ll} \max_{p(\mathbf{x})} & h(\bX)-\mu h(\bX+\bZ) \\
\mathrm{subject\;to} & \Cov(\bX) \preceq \bS, \end{array}
\label{opt:dcr2} \eeq where the maximization is over all random
vector $\bX$ independent of $\bZ$. For any $\mu \in \real$ and any
positive semidefinite $\bS$, a Gaussian $\bX$ is an optimal solution
of this optimization problem.
\end{corollary}

Observe that the optimization problem \eqref{opt:dcr2} is simply a
constrained version of the optimization problem \eqref{eq:Opt-UNC}.
Recall from Section \ref{sec:intro} that the optimization problem
\eqref{eq:Opt-UNC} can be solved by the classical EPI. Conversely,
it can be shown that the special case of the classical EPI with one
of the participant random vectors (say, $\bZ$ in \eqref{eq:EPI})
fixed to be Gaussian can also be obtained from the fact that a
Gaussian $\bX$ is an optimal solution of the optimization problem
\eqref{eq:Opt-UNC}. This can be done as follows. Choosing \beq \mu =
1+\exp\left[\frac{2}{n}\left(h(\bZ)-h(\bX)\right)\right], \eeq we
have from \eqref{eq:t} that \beq
h(\bX_G^*)=h(\bZ)-\frac{n}{2}\log(\mu-1)=h(\bX). \label{la1} \eeq
Since $\bX_G^*$ is an optimal solution of the optimization problem
\eqref{eq:Opt-UNC} (recall that $\bX_G^*$ has a special covariance
structure of being proportional to $\bK_Z$), we have \beq h(\bX)-\mu
h(\bX+\bZ) \leq h(\bX_G^*)-\mu h(\bX_G^*+\bZ). \label{la2} \eeq
Substituting \eqref{la1} into \eqref{la2}, we have $h(\bX+\bZ) \geq
h(\bX_G^*+\bZ)$ for any random vector $\bX$ independent of $\bZ$ and
satisfying $h(\bX)=h(\bX_G^*)$. This is precisely the Costa-Cover
form of the classical EPI \cite[Theorem~6]{DCT91}, so we have proved
the converse statement.

In light of the above statements, Corollary \ref{cor:dcr2} can be
thought of a generalization of the classical EPI. For technical
reasons, we were not able to prove Corollary \ref{cor:dcr2} directly
from Corollary \ref{cor:dcr} by letting $\bK_{Z_1}$ vanish. Instead,
we can resort to arguments (direct and perturbation ones) similar to
those for Theorem \ref{theorem:main} to prove Corollary
\ref{cor:dcr2}. Observe that in the optimization problem
\eqref{opt:dcr2} the lower constraint $\bK_X \succeq 0$ never bites,
so no enhancement is needed in the perturbation proof. The details
of the proof is omitted from the paper.

We now turn to the other degraded case where $\bK_{Z_1} \succeq
\bK_{Z_2}$. Consider the optimization problem
\beq \begin{array}{ll} \max_{p(\mathbf{x})} & h(\bX+\bZ_2+\bZ)-\mu h(\bX+\bZ_2) \\
\mathrm{subject\;to} & \Cov(\bX) \preceq \bS, \end{array}
\label{opt:dcw} \eeq where the maximization is over all random
vector $\bX$ independent of $\bZ_2$ and $\bZ$. For any $\mu \geq 1$,
by Theorem \ref{theorem:main} a Gaussian $\bX$ is an optimal
solution of this optimization problem. For $\mu \leq 0$, this is
also true because $h(\bX+\bZ_2+\bZ)$ and $h(\bX+\bZ_2)$ are
simultaneously maximized when $\bX$ is Gaussian with covariance
matrix $\bS$. However, as we shall see next, this is generally
\emph{not} the case for $\mu \in (0,1)$.

Consider the cases where \beq 0 \prec
\frac{\mu}{1-\mu}\bK_Z-\bK_{Z_2} \prec \bS. \label{cond} \eeq (Note
that this can only happen when $\mu \in (0,1)$ and also depends on
the realizations of $\bK_Z$, $\bK_{Z_2}$ and $\bS$.) Under this
assumption, we can verify that the covariance matrix $\bK_X^*$ of
$\bX_G^*$ must satisfy: \beq \bK_X^* =
\frac{\mu}{1-\mu}\bK_Z-\bK_{Z_2}. \eeq Let $\bX$ be a
\emph{non-Gaussian} random vector satisfying:
\begin{enumerate}
\item $h(\bX+\bZ_2)=h(\bX_G^*+\bZ_2)$; \item $\Cov(\bX) \preceq
\bS$.
\end{enumerate} Such an $\bX$ exists because by the assumption, $\bK_X^*$
is \emph{strictly} between $0$ and $\bS$. Since $\bX$ is
non-Gaussian, by the Costa-Cover form of the classical EPI, we have
\beq h(\bX+\bZ_2+\bZ)
> h(\bX_G^*+\bZ_2+\bZ). \eeq We thus
conclude that at least for the cases where the condition
\eqref{cond} holds, the optimal Gaussian solution $\bX_G^*$ cannot
be an optimal solution of the optimization problem \eqref{opt:dcw}.

\subsection{The Extremely-Skewed Case} Suppose that $\bZ_1$, $\bZ_2$ are in
$\real^2$. Let \beq \bK_{Z_1}=\bV_1\boldsymbol{\Sigma}_1\bV_1^t,
\quad \bK_{Z_2}=\bV_2\boldsymbol{\Sigma}_2\bV_2^t, \label{eq:eigen}
\eeq where $\bV_1$, $\bV_2$ are orthogonal matrices and \beq
\boldsymbol{\Sigma}_1=\Diag(\lambda_{11},\lambda_{12}), \quad
\boldsymbol{\Sigma}_2=\Diag(\lambda_{21},\lambda_{22}) \eeq are
diagonal matrices. Consider the limiting situation where
$\lambda_{12},\,\lambda_{21} \rightarrow \infty$, while
$\lambda_{11}$, $\lambda_{22}$ are kept fixed. Compared with the
degraded case where $\bK_{Z_1}$ dominates $\bK_{Z_2}$ in every
possible direction (or vice versa), this situation between
$\bK_{Z_1}$ and $\bK_{Z_2}$ is extremely skewed. We have the
following result.

\begin{corollary} \label{cor:es}
Let $Z$ be a Gaussian random variable, and let $\bv_1$, $\bv_2$ be
two deterministic vectors in $\real^2$. Consider the optimization
problem \beq
\begin{array}{ll}
\max_{p(\mathbf{x})} & h(\bv_1^t\bX+Z)-\mu h(\bv_2^t\bX+Z) \\
\mathrm{subject\;to} & \Cov(\bX) \preceq \bS, \end{array} \eeq where
the maximization is over all random vector $\bX$ (in $\real^2$)
independent of $Z$. For any $\mu \geq 1$ and any positive
semidefinite $\bS$, a Gaussian $\bX$ is an optimal solution of this
optimization problem.
\end{corollary}

\emph{Proof.} See Appendix \ref{app:es}. \hfill $\square$

Next, we use Corollary \ref{cor:es} to solve an optimization problem
that involves maximizing the differential entropy of the sum of two
\emph{dependent} random variables. To put it in perspective, let us
first consider the following simple optimization problem: \beq
\begin{array}{ll}
\max_{p(x_1,x_2)} & h(X_1+X_2) \\
\mathrm{subject\;to} & \Var(X_1) \leq a_1, \quad \Var(X_2) \leq a_2,
\end{array} \label{opt:Gaussian} \eeq where $a_1,a_2 \geq 0$ are
real numbers, $\Var(X)$ denotes the variance of $X$, and the
maximization is over all jointly distributed random variables
$(X_1,X_2)$. The solution to this optimization problem is clear:
$h(X_1+X_2)$ is maximized when $X_1$, $X_2$ are jointly Gaussian
with variance $a_1$ and $a_2$, respectively, and are aligned, i.e.,
$X_1=\sqrt{a_1/a_2}X_2$ almost surely.

Replacing both variance constraints in the optimization problem
\eqref{opt:Gaussian} by the entropy constraints, we have the
following optimization problem: \beq
\begin{array}{ll}
\max_{p(x_1,x_2)} & h(X_1+X_2) \\
\mathrm{subject\;to} & h(X_1) \leq a_1, \quad h(X_2) \leq a_2,
\end{array} \label{opt:CZ} \eeq where $a_1,a_2 \in \real$,
and the maximization is over all jointly distributed random
variables $(X_1,X_2)$. Different from the optimization problem
\eqref{opt:Gaussian}, a jointly Gaussian $(X_1,X_2)$ is \emph{not}
always an optimal solution of \eqref{opt:CZ}. This can seen as
follows. Consider the case $a_1=a_2$. Let $(X_{1G}^*,X_{2G}^*)$ be
the optimal Gaussian solution of the optimization problem
\eqref{opt:CZ}. We have $X_{1G}^*=X_{2G}^*$ almost surely, i.e.,
$X_{1G}^*$ and $X_{2G}^*$ are aligned and have the same marginal
distribution. Consider all jointly distributed random variables
$(X_1,X_2)$ for which $X_1$, $X_2$ have the same marginal density
function $f$ which satisfies:
\begin{enumerate}
\item $h(X_1)=h(X_{1G}^*)$; \item $f$ is \emph{not} log-concave.
\end{enumerate} The classical result of Cover and Zhang \cite{CZ94}
asserts that among all $(X_1,X_2)$ satisfying the above conditions,
there is at least one that satisfies \beq h(X_1+X_2)
> h(2X_1) = h(2X_{1G}^*) = h(X_{1G}^* + X_{2G}^*). \eeq
We thus conclude that a jointly Gaussian $(X_1,X_2)$ is not always
an optimal solution of the optimization problem \eqref{opt:CZ}.

Between \eqref{opt:Gaussian} and \eqref{opt:CZ} is the following
optimization problem: \beq
\begin{array}{ll}
\max_{p(x_1,x_2)} & h(X_1+X_2) \\
\mathrm{subject\;to} & \Var(X_1) \leq a_1, \quad h(X_2) \leq a_2,
\end{array} \label{opt:LV} \eeq where $a_1$, $a_2$ are real numbers
with $a_1 \geq 0$, and the maximization is over all jointly
distributed random variables $(X_1,X_2)$. The question whether a
Gaussian $(X_1,X_2)$ is an optimal solution of this optimization
problem remains, to our best knowledge, an open problem. The
following result, however, can be proved using Corollary
\ref{cor:es}.

\begin{corollary} \label{cor:LV}
Let $Z$ be a Gaussian variable, and let $a_1$, $a_2$ be real numbers
with $a_1 \geq 0$. Consider the optimization problem \beq
\begin{array}{ll}
\max_{p(x_1,x_2)} & h(X_1+X_2+Z) \\
\mathrm{subject\;to} & \Var(X_1) \leq a_1, \quad h(X_2+Z) \leq a_2,
\end{array} \label{app:LV2} \eeq where the maximization is over all
jointly distributed random variables $(X_1,X_2)$ independent of $Z$.
A Gaussian $(X_1,X_2)$ is an optimal solution of this optimization
problem for any $a_1 \geq 0$ and any $h(Z) \leq a_2 \leq a_2^*$
where \beq a_2^*=\frac{1}{2}\log\left(2\pi
e\left(\Var(Z)+\frac{1}{4}\left(\sqrt{a_1+4\Var(Z)}-\sqrt{a_1}\right)^2\right)\right).
\label{eq:a2} \eeq
\end{corollary}

\emph{Proof.} See Appendix \ref{app:LV}. \hfill $\square$

\section{Applications in Multiterminal Information Theory}
\label{sec:motiv}
\subsection{The Vector Gaussian Broadcast Channel}
We now use our main result to give an exact characterization of the
capacity region of the vector Gaussian broadcast channel. The
capacity region of the vector Gaussian broadcast channel was first
characterized by Weigarten et al. \cite{WSS05}.

Consider the following two-user discrete-time vector Gaussian
broadcast channel: \beq \bY_k[m] = \bX[m] + \bZ_k[m], \quad k=1,2,
\label{eq:AMBC} \eeq where $\{\bX[m]\}$ is the channel input subject
to an average matrix power constraint \beq
\frac{1}{N}\sum_{m=1}^{N}\bX[m]\bX^t[m] \preceq \bS,
\label{eq:Cons2} \eeq and the noise $\{\bZ_k[m]\}$ is i.i.d.
Gaussian with zero mean and strictly positive definite covariance
matrix $\bK_{Z_k}$ and is independent of $\{\bX[m]\}$. The
covariance structure of the Gaussian noise models a scalar Gaussian
broadcast channel with memory. Alternatively, it can also model the
downlink of a cellular system with multiple antennas; this was the
motivation of \cite{WSS05}.

A vector Gaussian broadcast is in general a \emph{nondegraded}
broadcast channel. An exact characterization of the capacity region
had been a long-standing open problem in multiterminal information
theory, particularly when viewed in the context of a scalar Gaussian
broadcast channel with memory. Prior to \cite{WSS05}, only bounds
were known. An outer bound, derived by Marton and K\"{o}rner
\cite[Theorem~5]{Marton79}, is given by $\mathcal{O}=\mathcal{O}_1
\cap \mathcal{O}_2$, where $\mathcal{O}_1$ is the set of rate pairs
$(R_1,R_2)$ satisfying \beqa
R_1 & \leq & I(\bX;\bY_1|U) \\
R_2 & \leq & I(U;\bY_2) \label{eq:Marton1} \eeqa for some
$p(\mathbf{y}_1,\mathbf{y}_2,\mathbf{x},u)=p(\mathbf{y}_1,\mathbf{y}_2|\mathbf{x})
p(\mathbf{x},u)$ such that $p(\mathbf{y}_1,\mathbf{y}_2|\mathbf{x})$
is the channel transition matrix and $p(\mathbf{x})$ satisfies the
constraint $\expect[\bX\bX^t] \preceq \bS$, and $\mathcal{O}_2$ is
the set of rate pairs $(R_1,R_2)$ satisfying \beqa
R_1 & \leq & I(V;\bY_1) \\
R_2 & \leq & I(\bX;\bY_2|V) \label{eq:Marton2} \eeqa for some
$p(\mathbf{y}_1,\mathbf{y}_2,\mathbf{x},v)=p(\mathbf{y}_1,\mathbf{y}_2|\mathbf{x})
p(\mathbf{x},v)$ such that $p(\mathbf{y}_1,\mathbf{y}_2|\mathbf{x})$
is the channel transition matrix and $p(\mathbf{x})$ satisfies the
constraint $\expect[\bX\bX^t] \preceq \bS$.

Next, we derive a tight upper bound on the achievable weighted sum
rate \beq \mu_1R_1+\mu_2R_2, \eeq using the Marton-K\"{o}rner outer
bound as the starting point. Since a capacity region is always
convex (per time-sharing argument), an exact characterization of all
the achievable weighted sum rates for all nonnegative $\mu_1,\mu_2$
provides an exact characterization of the entire capacity region.
First consider the case $\mu_2 \geq \mu_1 \geq 0$. By the
Marton-K\"{o}rner outer bound, any achievable rate pair $(R_1,R_2)$
must satisfy: \beqa \hspace{-22pt} \mu_1 R_1+\mu_2 R_2 & \leq &
\mu_1 \cdot \max\left\{I(\bX;\bY_1|U)+\mu{I}(U;\bY_2)\right\}
\label{LL}
\\ & = & \mu_1 \cdot
\max\left\{-h(\bZ_1)+\mu{h}(\bX+\bZ_2)+
\left[h(\bX+\bZ_1|U)-\mu{h}(\bX+\bZ_2|U)\right]\right\}. \label{LL0}
\eeqa Here $\mu=\frac{\mu_2}{\mu_1} \geq 1$, and the maximization is
over all $(U,\bX)$ independent of $(\bZ_1,\bZ_2)$ and satisfying the
matrix constraint $\expect[\bX\bX^t] \preceq \bS$. Consider the
terms $h(\bZ_1)$, $h(\bX+\bZ_2)$ and
$h(\bX+\bZ_1|U)-\mu{h}(\bX+\bZ_2|U)$ separately. We have \beq
h(\bZ_1)=\frac{1}{2}\log\left((2\pi{e})^n|\bK_{Z_1}|\right)
\label{LL1} \eeq and \beq h(\bX+\bZ_2) \leq
\frac{1}{2}\log\left((2\pi{e})^n|\bS+\bK_{Z_2}|\right). \label{LL2}
\eeq Further note that maximizing $h(\bX+\bZ_1|U)-\mu
h(\bX+\bZ_2|U)$ is simply a conditional version of the optimization
problem \eqref{eq:Opt-G}. We have the following result, which is a
conditional version of Theorem \ref{theorem:main}.

\begin{theorem} \label{thm:con}
Let $\bZ_1$, $\bZ_2$ be two Gaussian vectors with strictly positive
definite covariance matrices $\bK_{Z_1}$ and $\bK_{Z_2}$,
respectively. Let $\mu \geq 1$ be a real number, $\bS$ be a positive
semidefinite matrix, and $U$ be a random variable independent of
$\bZ_1$ and $\bZ_2$. Consider the optimization problem
\beq \begin{array}{ll} \max_{p(\mathbf{x}|u)} & h(\bX+\bZ_1|U)-\mu h(\bX+\bZ_2|U) \\
\mathrm{subject\;to} & \Cov(\bX|U) \preceq \bS, \end{array}
\label{eq:Opt-GA} \eeq where the maximization is over all
conditional distribution of $\bX$ given $U$ independent of $\bZ_1$
and $\bZ_2$. A Gaussian $p(\mathbf{x}|u)$ with the same covariance
matrix for each $u$ is an optimal solution of this optimization
problem.
\end{theorem}

The result of the above theorem has two parts. The part that says a
Gaussian $p(\mathbf{x}|u)$ is an optimal solution follows directly
from Theorem \ref{theorem:main}; the part that says the optimal
Gaussian $p(\mathbf{x}|u)$ has the same covariance matrix for each
$u$ is equivalent to that the optimal value of the optimization
problem \eqref{eq:Opt-G} is a \emph{concave} function of $\bS$.
Despite being a matrix problem, a direct proof of the concavity
turns out to be difficult. Instead, Theorem \ref{thm:con} can be
proved following the same footsteps as those for Theorem
\ref{theorem:main}, except that we need to replace the classical EPI
by a conditional version proved by Bergmans \cite[Lemma
II]{Bergman74}. Let $\bZ$ be a Gaussian vector. Bergmans'
conditional EPI states that \beq
\exp\left[\frac{2}{n}h(\bX+\bZ|U)\right] \geq
\exp\left[\frac{2}{n}h(\bX|U)\right] +
\exp\left[\frac{2}{n}h(\bZ)\right] \label{eq:BEPI} \eeq for any
$(\bX,U)$ independent of $\bZ$. The equality holds if and only if
conditional on $U=u$, $\bX$ is Gaussian with a covariance matrix
proportional to that of $\bZ$ and has the same covariance matrix for
each $u$. The details of the proof are omitted from the paper.

By Theorem \ref{thm:con}, we have \beq
h(\bX+\bZ_1|U)-\mu{h}(\bX+\bZ_2|U) \leq \max_{0 \preceq \bK_X
\preceq
\bS}\left\{\frac{1}{2}\log\left((2\pi{e})^n|\bK_X+\bK_{Z_1}|\right)-
\frac{\mu}{2}\log\left((2\pi{e})^n|\bK_X+\bK_{Z_2}|\right)\right\}.
\label{LL3} \eeq Substituting \eqref{LL1}, \eqref{LL2} and
\eqref{LL3} into \eqref{LL0}, we obtain \beq \mu_1R_1+\mu_2R_2 \leq
\max_{0 \preceq \bK_X \preceq
\bS}\left\{\frac{\mu_1}{2}\log\left|\frac{\bK_X+\bK_{Z_1}}{\bK_{Z_1}}\right|+\frac{\mu_2}{2}
\log\left|\frac{\bS+\bK_{Z_2}}{\bK_X+\bK_{Z_2}}\right|\right\}.
\label{LL5} \eeq Note that the weighted sum rates give by
\eqref{LL5} can be achieved by \emph{dirty-paper coding}
\cite{CS03,YSJCC01}, so \eqref{LL5} is an \emph{exact}
characterization of all the achievable weighted sum rates for $\mu_2
\geq \mu_1 \geq 0$.

For $\mu_1 \geq \mu_2 \geq 0$, we have from the Marton-K\"{o}rner
bound that \beq \mu_1R_1+\mu_2R_2 \leq \mu_2\cdot
\max_{p(\mathbf{X},V)}\left\{\mu{I}(V;\bY_1)+
I(\bX;\bY_2|V)\right\}. \eeq Here $\mu = \frac{\mu_1}{\mu_2} \geq
1$, and the maximization is over all $(V,\bX)$ independent of
$(\bZ_1,\bZ_2)$ and satisfying the matrix constraint
$\expect[\bX\bX^t] \preceq \bS$. Relabeling $V$ as $U$, the
optimization problem becomes identical to that in \eqref{LL}. We
thus conclude that \beq \mu_1R_1+\mu_2R_2 \leq \max_{0 \preceq \bK_X
\preceq
\bS}\left\{\frac{\mu_1}{2}\log\left|\frac{\bS+\bK_{Z_1}}{\bK_X+\bK_{Z_1}}\right|
+\frac{\mu_2}{2}\log\left|\frac{\bK_X+\bK_{Z_2}}{\bK_{Z_2}}\right|\right\}.
\eeq is an exact characterization of all the achievable weighted sum
rates for $\mu_1 \geq \mu_2 \geq 0$. This settles the problem of
characterizing the entire capacity region of the vector Gaussian
broadcast channel.

\subsection{Distributed Source Coding with a Single Quadratic Distortion Constraint}
Our result is also relevant in the following distributed source
coding problem. Let $\{\bY_1[m]\}$, $\{\bY_2[m]\}$ be two i.i.d.
vector Gaussian sources with strictly positive definite covariance
matrix $\bK_{Y_1}$ and $\bK_{Y_2}$, respectively. At each time $m$,
$\bY_1[m]$ and $\bY_2[m]$ are jointly Gaussian. The encoder is only
allowed to perform \emph{separate} encoding on the sources. The
decoder, on the other hand, can reconstruct the sources based on
\emph{both} encoded messages. We wish to characterize the entire
rate region for which the quadratic distortion for reconstructing
$\{\bY_1[m]\}$ at the decoder \beq
\frac{1}{N}\sum_{m=1}^{N}\left(\bY_1[m]-\widehat{\bY}_1[m]\right)
\left(\bY_1[m]-\widehat{\bY}_1[m]\right)^t \preceq \bD. \eeq (There
is no distortion constraint on the source $\{\bY_2[m]\}$.) This is
the so-called \emph{distributed source coding with a single
quadratic distortion constraint problem}.

Note that $\bY_1[m]$, $\bY_2[m]$ are jointly Gaussian, so without
loss of generality we can write \beq \bY_1[m] = \bA\bY_2[m]+\bZ[m],
\eeq where $\bA$ is an invertible matrix and $\bZ[m]$ is Gaussian
and independent of $\bY_2[m]$. Since there is no distortion
constraint on $\{\bY_2[m]\}$, we can always assume that $\bY_1[m]$
is a degraded version of $\bY_2[m]$ by relabeling $\bA\bY_2[m]$ as
$\bY_2[m]$. In this case, an outer bound can be obtained similarly
to that for the discrete memoryless degraded broadcast channel
\cite{Gallager74}: \beq \begin{array}{lll}
R_1 & \geq & I(\bY_1;\widehat{\bY}_1|U) \\
R_2 & \geq & I(U;\bY_2)
\end{array} \label{eq:SW} \eeq
for some $p(u,\widehat{\mathbf{y}}_1,\mathbf{y}_1,\mathbf{y}_2)
=p(\widehat{\mathbf{y}}_1|u,\mathbf{y}_1)p(u|\mathbf{y}_2)p(\mathbf{y}_1,\mathbf{y}_2)$,
where $p(\mathbf{y}_1,\mathbf{y}_2)$ is the joint distribution of
the sources and $p(\widehat{\mathbf{y}}_1|u,\mathbf{y}_1)$ satisfies
the matrix constraint
$\expect[(\bY_1-\widehat{\bY}_1)(\bY_1-\widehat{\bY}_1)^t] \preceq
\bD$. The proof is deferred to Appendix \ref{app:CEO}. Next, we
derive a lower bound on all the achievable weighted sum rates
$\mu_1R_1+\mu_2R_2$ for all nonnegative $\mu_1, \mu_2$, using this
outer bound as the starting point.

By the outer bound \eqref{eq:SW}, all the achievable rate pairs
$(R_1,R_2)$ must satisfy: \beqa \hspace{-30pt} \mu_1R_1+\mu_2R_2 &
\geq & \mu_1
\cdot\min_{p(u,\widehat{\mathbf{y}}|\mathbf{y}_1,\mathbf{y}_2)}\left\{
\mu I(\bY_1;\widehat{\bY}_1|U)+I(U;\bY_2)\right\}
\\ & = & \mu_1 \cdot \min_{p(u,\widehat{\mathbf{y}}|\mathbf{y}_1,\mathbf{y}_2)}\left\{
h(\bY_2)-\mu h(\bY_1|\widehat{\bY}_1,U)-\left[h(\bY_2|U)-\mu
h(\bY_2+\bZ|U)\right]\right\}. \label{TT} \eeqa Here
$\mu=\frac{\mu_2}{\mu_1} \geq 0$, and the minimization is over all
$p(u,\widehat{\mathbf{y}}|\mathbf{y}_1,\mathbf{y}_2)$ such that $U$
is independent of $\bZ$ and
$\expect[(\bY_1-\widehat{\bY}_1)(\bY_1-\widehat{\bY}_1)^t] \preceq
\bD$ is satisfied. Consider the terms $h(\bY_2)$,
$h(\bY_1|\widehat{\bY}_1,U)$ and $h(\bY_2|U)-\mu h(\bY_2+\bZ|U)$
separately. We have \beq h(\bY_2) =
\frac{1}{2}\log\left((2\pi{e})^n|\bK_{Y_2}|\right) \label{TT1} \eeq
and \beq h(\bY_1|\widehat{\bY}_1,U) =
h(\bY_1-\widehat{\bY}_1|\widehat{\bY}_1,U) \leq
h(\bY_1-\widehat{\bY}_1) \leq
\frac{1}{2}\log\left((2\pi{e})^n|\bD|\right). \label{TT2} \eeq Hence
we only need to maximize $h(\bY_2|U)-\mu h(\bY_2+\bZ|U)$ subject to
the constraints \beq \Cov(\bY_2|U) \preceq \bK_{Y_2} \quad
\mbox{and} \quad \Cov(\bY_1|U) \succeq \mathbf{D}.
\label{eq:constraints} \eeq In case that the constraint
$\Cov(\bY_1|U) \succeq \mathbf{D}$ does not bite, we can use (a
conditional version of) Corollary \ref{cor:dcr2} to show that a
Gaussian test channel between $\bY_2$ and $\bU$ is a maximizer: \beq
h(\bY_2|U)-\mu h(\bY_2+\bZ|U) \leq \max_{0 \preceq \bK \preceq
\bK_{Y_2}}\left\{\frac{1}{2}\log\left((2\pi{e})^n|\bK|\right)-
\frac{\mu}{2}\log\left((2\pi{e})^n|\bK+\bK_{Y_1}-\bK_{Y_2}|\right)\right\}.
\label{TT3} \eeq Substituting \eqref{TT1}, \eqref{TT2} and
\eqref{TT3} into \eqref{TT}, we have \beq \mu_1R_1+\mu_2R_2 \geq
\max_{0 \preceq \bK \preceq
\bK_{Y_2}}\left\{\frac{\mu_1}{2}\log\left|\frac{\bK_{Y_2}}{\bK}\right|+\frac{\mu_2}{2}
\log\left|\frac{\bK+\bK_{Y_1}-\bK_{Y_2}}{\bD}\right|\right\}. \eeq
On the other hand, this weighted sum rate can be achieved by the
following natural Gaussian separation scheme:
\begin{enumerate} \item Quantize $\{\bY_1[m]\}$ and
$\{\bY_2[m]\}$ separately using Gaussian codebooks; \item Use
\emph{Slepian-Wolf coding} \cite{SW73} on the quantized version of
$\{\bY_1[m]\}$, treating the quantized version of $\{\bY_2[m]\}$ as
decoder side information.
\end{enumerate} This would have settled the rate region for the
distributed source coding with a single quadratic constraint
problem.

Unfortunately, there are indeed instances where the constraint
$\Cov(\bY_1|U) \succeq \mathbf{D}$ cannot be ignored; in such cases,
the outer bound studied here will be strictly inside the inner bound
achieved by the natural Gaussian separation scheme.

\section{Concluding Remarks} \label{sec:con}
The classical EPI is an important inequality with interesting
connections to statistical estimation theory. In information theory,
it has been key to the proof of the converse coding theorem in
several important scalar Gaussian multiterminal problems
\cite{Bergman74,Oohama98,Ozarow80}. In the vector situation, the
equality condition of the classical EPI is stringent: the equality
requires the participating random vectors not only be Gaussian but
also have proportional covariance matrices. In several instances,
this coupling between the Gaussianity and the proportionality is the
main cause that prevents the classical EPI from being directly
useful in extending the converse proof from the scalar case to the
vector situation.

In this paper, we proved a new extremal inequality involving
entropies of random vectors. In one special case, this inequality
can be seen as a \emph{robust} version of the classical EPI. By
``robust", we refer to the fact that in the new extremal inequality,
the optimality of a Gaussian distribution does not couple with a
specific covariance structure, i.e. proportionality. We show that
the new extremal inequality is useful in evaluating certain
genie-aided outer bounds for the capacity/rate region for the vector
Gaussian broadcast channel and the distributed source coding with a
single quadratic constraint problems.

We offered two proofs to the new extremal inequality: one by
appropriately using the classical EPI, and the other by the
perturbation approach of Stam \cite{Stam59} and Blachman
\cite{Blachman65}. The perturbation approach gives more insights: it
takes the problem (via the de Bruijn identity) to the Fisher
information domain where the proportionality no longer seems a
hurdle. Whereas the advantage of the perturbation approach is not
crucial for the entropy inequalities discussed in this paper, it
becomes crucial in some other situations \cite{WLVSS06} where the
enhancement technique of Weingarten et al. does not suffice.

\begin{appendix}
\section{A Direct Proof of Theorem \ref{theorem:main}}
\label{app:basicproof} We now show that the classical EPI can be
appropriately used to prove Theorem \ref{theorem:main}. We first
give the outline of the proof.

\emph{Proof Outline.} We first show that without loss of generality,
we can assume that $\bS$ is \emph{strictly} positive definite. Next,
we denote the optimization problem \eqref{eq:Opt-G} by $P$ and the
optimal value of $P$ by $(P)$. To show that a Gaussian $\bX$ is an
optimal solution of $\mathrm{P}$, it is sufficient to show that
$(\mathrm{P})=(P_G)$, where $P_G$ is the Gaussian version of $P$ by
restricting the solution space within Gaussian distributions: \beq
\begin{array}{ll} \max_{\bK_X} &
\frac{1}{2}\log\left((2\pi{e})^n\left|\bK_X+\bK_{Z_1}\right|\right)
-\frac{\mu}{2}\log\left((2\pi{e})^n\left|\bK_X+\bK_{Z_2}\right|\right) \\
\mbox{subject to} & \mathbf{0} \preceq \bK_X \preceq \bS.
\end{array} \label{MT} \eeq Since restricting the solution space
can only decrease the optimal value of a maximization problem, we
readily have $(P) \geq (P_G)$. To prove the reverse inequality $(P)
\leq (P_G)$, we shall consider an auxiliary optimization problem
$\widetilde{P}$ and its Gaussian version $\widetilde{P}_G$. In
particular, we shall construct a $\widetilde{P}$ such that: \beq (P)
\leq (\widetilde{P}), \quad (\widetilde{P})=(\widetilde{P}_G), \quad
(\widetilde{P}_G)=(P_G). \label{eq:A0} \eeq We will then have $(P)
\leq (P_G)$ and hence $(P)=(P_G)$.

The proof is rather long, so we divide it into several steps.

\emph{Step 1: $\bS \succeq 0$, $|\bS|=0$.} We show that for any $\bS
\succeq 0$ but $|\bS|=0$, there is an \emph{equivalent} optimization
problem of type \eqref{eq:Opt-G} in which the the upper bound on
$\bX$ is strictly positive definite.

Suppose that the rank of $\bS$ is $r<n$, i.e., $\bS$ is rank
deficient. Let \beq \bS = \bQ_S\boldsymbol{\Sigma}_S\bQ_S^t, \eeq
where $\bQ_S$ is an orthogonal matrix, and \beq
\boldsymbol{\Sigma}_S = \Diag(\lambda_1,\cdots,\lambda_r,0,\cdots,0)
\eeq is a diagonal matrix. For any $\bX \preceq \bS$, let
$\overline{\bX}=\left(\overline{\bX}_a^t,\overline{\bX}_b^t\right)^t=\bQ_S^t\bX$
where $\overline{\bX}_a$ is of a length $r$. We have \beq
\Cov(\overline{\bX}) = \bQ_S^t\Cov(\bX)\bQ_S \preceq \bQ_S^t\bS\bQ_S
= \boldsymbol{\Sigma}_S, \eeq which implies that
$\Cov(\overline{\bX}_b)=0$, i.e., $\overline{\bX}_b$ is
deterministic. Without loss of generality, let us assume that
$\overline{\bX}_b=0$. So an optimization over $\Cov(\bX) \preceq
\bS$ is the same as an optimization over $\Cov(\overline{\bX}_a)
\preceq \Diag(\lambda_1,\cdots,\lambda_r)$.

Next, let \beq \bQ_S^t\bK_{Z_i}\bQ_S = \left(
                                     \begin{array}{cc}
                                       \bA_i & \bB_i^t \\
                                       \bB_i & \bC_i \\
                                     \end{array}
                                   \right) \eeq where
$\bA_i$, $\bB_i$ and $\bC_i$ are submatrices of size $r \times r$,
$(n-r) \times r$, and $(n-r) \times (n-r)$, respectively, and let
\beq \bD_i = \left(
               \begin{array}{cc}
                 \bI & -\bB_i^t\bC_i^{-1} \\
                 0 & \bI \\
               \end{array}
             \right). \eeq
We have \beq \bD\bQ_S^t\bX = \left(
               \begin{array}{cc}
                 \bI & -\bB_i^t\bC_i^{-1} \\
                 0 & \bI \\
               \end{array}
             \right) \left(
                       \begin{array}{c}
                         \overline{\bX}_a \\
                         0 \\
                       \end{array}
                     \right) = \left(
                       \begin{array}{c}
                         \overline{\bX}_a \\
                         0 \\
                       \end{array}
                     \right), \eeq
and \beq \Cov(\bD\bQ_S^t\bZ_i) = \left(
               \begin{array}{cc}
                 \bI & -\bB_i^t\bC_i^{-1} \\
                 0 & \bI \\
               \end{array}
             \right) \left(
                                     \begin{array}{cc}
                                       \bA_i & \bB_i^t \\
                                       \bB_i & \bC_i \\
                                     \end{array}
                                   \right) \left(
               \begin{array}{cc}
                 \bI & 0  \\
                 -\bC_i^{-1}\bB_i & \bI \\
               \end{array}
             \right) = \left(
                         \begin{array}{cc}
                           \bA_i-\bB_i^t\bC_i^{-1}\bB_i & 0 \\
                           0 & \bC_i \\
                         \end{array}
                       \right).
             \eeq
Hence if we let $\bD\bQ_S^t\bZ_i =
(\overline{\bZ}_{i,a}^t,\overline{\bZ}_{i,b}^t)^t$ where
$\overline{\bZ}_{i,a}$ is of a length $r$, then
$\overline{\bZ}_{i,a}$ and $\overline{\bZ}_{i,b}$ are statistically
independent. It follows that \beq
h(\bX+\bZ_i)=h(\bD\bQ_S^t\bX+\bD\bQ_S^t\bZ_i)=
h(\bX_a+\overline{\bZ}_{i,a},\overline{\bZ}_{i,b}) =
h(\bX_a+\overline{\bZ}_{i,a})+h(\overline{\bZ}_{i,b}). \eeq So
maximizing $h(\bX+\bZ_1)-\mu h(\bX+\bZ_2)$ is equivalent to
maximizing $h(\overline{\bX}_a+\overline{\bZ}_{1,a})-\mu
h(\overline{\bX}_a+\overline{\bZ}_{2,a})
+h(\overline{\bZ}_{1,b})-\mu h(\overline{\bZ}_{2,b})$. Note that
$h(\overline{\bZ}_{i,b})$, $i=1,2$, are constants. Hence to show
that \eqref{eq:Opt-G} has a Gaussian optimal solution for a rank
deficient $\bS$, it is sufficient to show that \beq
\begin{array}{ll} \max_{p(\mathbf{x_a})} &
h(\overline{\bX}_a+\overline{\bZ}_{1,a})-h(\overline{\bX}_a+\overline{\bZ}_{1,a}) \\
\mathrm{subject\;to} & \Cov(\overline{\bX}_a) \preceq
\Diag(\lambda_1,\cdots,\lambda_r),
\end{array} \eeq has a Gaussian optimal solution. Since $\Diag(\lambda_1,\cdots,\lambda_r)$
now has a full rank, we conclude that without loss of generality, we
may assume that $\bS$ in \eqref{eq:Opt-G} is strictly positive
definite.

\emph{Step 2: Construction of $\widetilde{P}$.} Let $\bX_G^*$ be an
optimal Gaussian solution of $P$, and let $\bK_X^*$ be the
covariance matrix of $\bX_G^*$. Then $\bK_X^*$ is an optimal
solution to the optimization problem \eqref{MT}. Although this conic
program is generally nonconvex, it was shown in
\cite[Lemma~5]{WSS05} that for $\bS \succ 0$, $\bK_X^*$ must satisfy
the following KKT-like conditions: \beqa
\frac{1}{2}(\bK_X^*+\bK_{Z_1})^{-1}+\bM_1 & =&
\frac{\mu}{2}(\bK_X^*+\bK_{Z_2})^{-1}+\bM_2 \label{eq:A1} \\
\bM_1 \bK_X^* & = & 0 \label{eq:A2} \\
\bM_2(\bS-\bK_X^*) & = & 0, \label{eq:A3} \eeqa where $\bM_1, \bM_2
\succeq 0$ are Lagrange multipliers corresponding to $\bK_X \succeq
0$ and $\bK_X \preceq \bS$, respectively. Let
$\bK_{\widetilde{Z}_1}$, $\bK_{\widetilde{Z}_2}$ be two real
symmetric matrices satisfying \beqa
\frac{1}{2}(\bK_X^*+\bK_{Z_1})^{-1}+\bM_1 & = &
\frac{1}{2}(\bK_X^*+\bK_{\widetilde{Z}_1})^{-1},
\label{eq:A4} \\
\frac{\mu}{2}(\bK_X^*+\bK_{Z_2})^{-1}+\bM_2 & = &
\frac{\mu}{2}(\bK_X^*+\bK_{\widetilde{Z}_2})^{-1}. \label{eq:A5}
\eeqa We have the following results on $\bK_{\widetilde{Z}_1}$ and
$\bK_{\widetilde{Z}_2}$ proved in \cite[Lemma~11,12]{WSS05}.
\begin{lemma} \label{lemma:eh1}
For $\bK_X^*$, $\bK_{Z_i}$, $\bK_{\widetilde{Z}_i}$, $\bM_i$,
$i=1,2$, related through \eqref{eq:A1} to \eqref{eq:A5}, and $\mu
\geq 1$, we have \beqa 0 \;\; \preceq &
\bK_{\widetilde{Z}_1} & \preceq \;\; \bK_{Z_1}, \\
\bK_{\widetilde{Z}_1} \;\; \preceq & \bK_{\widetilde{Z}_2} & \preceq
\;\; \bK_{Z_2}. \eeqa
\end{lemma}
The matrices $\bK_{\widetilde{Z}_1}$, $\bK_{\widetilde{Z}_2}$ are
positive semidefinite, so they can serve as covariance matrices. Let
$\widetilde{\bZ}_1$, $\widetilde{\bZ}_2$ be two Gaussian vectors
with covariance matrix $\bK_{\widetilde{Z}_1}$ and
$\bK_{\widetilde{Z}_2}$, respectively. Let us define the auxiliary
optimization problem $\widetilde{P}$ as: \beq
\begin{array}{ll} \max_{p(\mathbf{x})} & h(\bX+\widetilde{\bZ}_1)-\mu
h(\bX+\widetilde{\bZ}_2)+F \\
\mbox{subject to} & \Cov(\bX) \preceq \bS, \end{array}
\label{opt:aux3} \eeq where the constant \beq F :=
h(\bZ_1)-h(\widetilde{\bZ}_1) +
\mu\left(h(\bX_G^{(S)}+\widetilde{\bZ}_2)-h(\bX_G^{(S)}+\bZ_2)\right),
\eeq $\bX_G^{(S)}$ is a Gaussian vector with covariance matrix $\bS$
and independent of $\bZ_2$ and $\widetilde{\bZ}_2$, and the
maximization is over all random vector $\bX$ independent of
$\widetilde{\bZ}_1$ and $\widetilde{\bZ}_2$.

In \cite[p.~3937]{WSS05}, the authors call the process of replacing
$\bZ_1$ and $\bZ_2$ with $\widetilde{\bZ}_1$ and
$\widetilde{\bZ}_2$, respectively, \emph{enhancement}. Next, we show
that the auxiliary optimization problem $\widetilde{P}$ defined in
\eqref{opt:aux3} satisfies the desired chain of relationships
\eqref{eq:A0}.

\emph{Step 3: Proof of $(P)\leq(\widetilde{P})$.} Note that
$\mathrm{P}$ and $\widetilde{\mathrm{P}}$ have the same solution
space. So to show that $(P) \leq (\widetilde{P})$, it is sufficient
to show that for each admissible solution, the value of the
objective function of $P$ is less or equal to that of
$\widetilde{P}$.

The difference between the objective functions of $\mathrm{P}$ and
$\widetilde{\mathrm{P}}$ can be written as \beqa \hspace{-40pt} & &
h(\bX+\bZ_1)-h(\bZ_1) -h(\bX+\widetilde{\bZ}_1)+h(\widetilde{\bZ}_1)
\nonumber
\\ \hspace{-40pt} & & \hspace{100pt} -\,\mu\left(h(\bX+\bZ_2)-h(\bX+\widetilde{\bZ}_2) -
h(\bX_S+\bZ_2)+h(\bX_S+\widetilde{\bZ}_2)\right). \label{eq:A6}
\eeqa By Lemma \ref{lemma:eh1}, $\bK_{Z_i} \succeq
\bK_{\widetilde{Z}_i}$ for $i=1,2$. So we can write
$\bZ_i=\widetilde{\bZ}_i+\widehat{\bZ}_i$, where $\widehat{\bZ}_i$
is a Gaussian vector independent of $\widetilde{\bZ}_i$. We have
\beqa \hspace{-20pt}
h(\bX+\bZ_1)-h(\bZ_1)-h(\bX+\widetilde{\bZ}_1)+h(\widetilde{\bZ}_1)
& = & I(\bX;\bX+\bZ_1) - I(\bX;\bX+\widetilde{\bZ}_1) \label{ww}\\
& = & I(\bX;\bX+\widetilde{\bZ}_1+\widehat{\bZ}_1)-
I(\bX;\bX+\widetilde{\bZ}_1) \\ & \leq & 0, \label{eq:A7} \eeqa
where the inequality is due to the Markov chain \beq \bX \rightarrow
\bX+\widetilde{\bZ}_1 \rightarrow
\bX+\widetilde{\bZ}_1+\widehat{\bZ}_1. \label{ww2} \eeq Further, let
$\bX_G$ be a Gaussian random vector with the same covariance matrix
as that of $\bX$. Assume that $\bX_G$ is independent of
$\widetilde{\bZ}_2$ and $\widehat{\bZ}_2$. Note that both $\bX_G$
and $\bX_G^{(S)}$ are Gaussian and that \beq \Cov(\bX_G) = \Cov(\bX)
\preceq \bS = \Cov(\bX_G^{(S)}). \eeq So we can write
$\bX_G^{(S)}=\bX_G+\widehat{\bX}_G$, where $\widehat{\bX}_G$ is a
Gaussian random vector independent of $\bX_G$. We have \beqa
\hspace{-20pt} & & h(\bX+\bZ_2)-h(\bX+\widetilde{\bZ}_2) -
h(\bX_G^{(S)}+\bZ_2)+h(\bX_G^{(S)}+\widetilde{\bZ}_2) \nonumber \\
\hspace{-20pt} & & \hspace{20pt} = \;
h(\bX+\widetilde{\bZ}_2+\widehat{\bZ}_2)-h(\bX+\widetilde{\bZ}_2)
-(h(\bX_G^{(S)}+\widetilde{\bZ}_2+\widehat{\bZ}_2)-h(\bX_G^{(S)}+\widetilde{\bZ}_2))
\\ \hspace{-20pt} & & \hspace{20pt} = \;
I(\widehat{\bZ}_2;\bX+\widetilde{\bZ}_2+\widehat{\bZ}_2) -
I(\widehat{\bZ}_2;\bX_G^{(S)}+\widetilde{\bZ}_2+\widehat{\bZ}_2)
\label{eq:A8} \\ \hspace{-20pt} & & \hspace{20pt} \geq \;
I(\widehat{\bZ}_2;\bX_G+\widetilde{\bZ}_2+\widehat{\bZ}_2) -
I(\widehat{\bZ}_2;\bX_G^{(S)}+\widetilde{\bZ}_2+\widehat{\bZ}_2)
\label{eq:A85} \\ \hspace{-20pt} & & \hspace{20pt} = \;
I(\widehat{\bZ}_2;\bX_G+\widetilde{\bZ}_2+\widehat{\bZ}_2) -
I(\widehat{\bZ}_2;\widehat{\bX}_G+\bX_G+\widetilde{\bZ}_2+\widehat{\bZ}_2)
\\ \hspace{-20pt} & & \hspace{20pt} \geq \; 0, \label{eq:A86}
\eeqa where inequality \eqref{eq:A85} follows from \beq
I(\widehat{\bZ}_2;\bX+\widetilde{\bZ}_2+\widehat{\bZ}_2) \geq
I(\widehat{\bZ}_2;\bX_G+\widetilde{\bZ}_2+\widehat{\bZ}_2) \eeq
which is due to the worst noise result of Lemma \ref{lemma:worst},
and inequality \eqref{eq:A86} follows from the Markov chain \beq
\widehat{\bZ}_2 \rightarrow \bX_G+\widetilde{\bZ}_2+\widehat{\bZ}_2
\rightarrow \widehat{\bX}_G+\bX_G+\widetilde{\bZ}_2+\widehat{\bZ}_2.
\eeq Substituting \eqref{eq:A7} and \eqref{eq:A86} into
\eqref{eq:A6}, we conclude that the difference between the objective
functions of $\mathrm{P}$ and $\widetilde{\mathrm{P}}$ is
nonpositive for any admissible $\bX$ (i.e., $\Cov(\bX) \preceq \bS$)
and any $\mu>1$.

\emph{Step 4: Proof of $(\widetilde{P})=(\widetilde{P}_G)$.} To show
that $(\widetilde{P})=(\widetilde{P}_G)$, it is sufficient to show
that $\bX_G^*$, the optimal solution of $P_G$, is also an optimal
solution of $\widetilde{P}$. We consider the cases $\mu=1$ and $\mu
> 1$ separately.

First assume that $\mu > 1$. By Lemma \ref{lemma:eh1},
$\bK_{\widetilde{Z}_2} \succeq \bK_{\widetilde{Z}_1}$. So we can
write $\widetilde{\bZ}_2 = \widetilde{\bZ}_1+\widetilde{\bZ}$, where
$\widetilde{\bZ}$ is Gaussian and independent of
$\widetilde{\bZ}_1$. We have \beqa \hspace{-25pt}
h(\bX+\widetilde{\bZ}_1)-\mu h(\bX+\widetilde{\bZ}_2) & = &
h(\bX+\widetilde{\bZ}_1)-\mu h(\bX+\widetilde{\bZ}_1+\widetilde{\bZ}) \\
& \leq & h(\bX+\widetilde{\bZ}_1)-\frac{\mu n}{2}
\log\left(\exp\left[\frac{2}{n}h(\bX+\widetilde{\bZ}_1)\right]+\exp\left[
\frac{2}{n}h(\widetilde{\bZ})\right]\right)
\label{eq:A12} \\
& \leq &
f\left(h(\widetilde{\bZ})-\frac{n}{2}\log(\mu-1);h(\widetilde{\bZ})\right),
\label{eq:A14} \eeqa where \eqref{eq:A12} follows from the classical
EPI, and the function $f$ in \eqref{eq:A14} was defined in
\eqref{eq:f}. Next, we verify that the upper bound on the right-hand
side of \eqref{eq:A14} is achieved by $\bX_G^*$. Substituting
\eqref{eq:A4} and \eqref{eq:A5} into the KKT-like condition
\eqref{eq:A1}, we obtain \beq (\bK_X^*+\bK_{\widetilde{Z}_1})^{-1} =
\mu(\bK_X^*+\bK_{\widetilde{Z}_2})^{-1}, \label{eq:AA} \eeq which
gives \beq \bK_X^*+\bK_{\widetilde{Z}_1} =
(\mu-1)^{-1}\bK_{\widetilde{Z}}. \label{eq:A} \eeq Hence,
$\bX_G^*+\widetilde{\bZ}_1$ and $\widetilde{\bZ}$ have proportional
covariance matrices and inequality \eqref{eq:A12} holds with
equality. Further by \eqref{eq:A}, \beq h(\bX_G^*+\widetilde{\bZ}_1)
= h(\widetilde{\bZ})-\frac{n}{2}\log(\mu-1). \label{eq:A15} \eeq A
comparison of \eqref{eq:A15} and \eqref{eq:t} confirms that
$h(\bX_G^*+\widetilde{\bZ}_1)$ achieves the global maxima of
function $f(t;h(\widetilde{\bZ}))$, i.e., inequality \eqref{eq:A14}
becomes equality with $\bX_G^*$. We thus conclude that $\bX_G^*$ is
an optimal solution of $\widetilde{P}$ for all $\mu>1$.

For $\mu=1$, we have from \eqref{eq:AA} that
$\bK_{\widetilde{Z}_1}=\bK_{\widetilde{Z}_2}$. So the objective
function of $\widetilde{P}$ is constant, and $\bX_G^*$ is trivially
an optimal solution of $\widetilde{P}$.

\emph{Step 5: Proof of $(\widetilde{P}_G)=(P_G)$.} Note that
$\bX_G^*$ is an optimal solution of both $\widetilde{P}_G$ and
$P_G$. So to show that $(\widetilde{P}_G)=(P_G)$, we only need to
compare the objective functions of $\widetilde{P}_G$ and $P_G$
evaluated at $\bX_G^*$. The following result, which is a minor
generalization of \cite[Lemma~11,12]{WSS05}, shows that the
objective functions of $\widetilde{P}_G$ and $P_G$ take equal values
at $\bX_G^*$.

\begin{lemma} \label{lemma:eh2}
For $\bK_X^*$, $\bK_{Z_i}$, $\bK_{\widetilde{Z}_i}$, $\bM_i$,
$i=1,2$, defined through \eqref{eq:A1} to \eqref{eq:A5} and $\mu
\geq 1$, we have \beqa
(\bK_X^*+\bK_{\widetilde{Z}_1})^{-1}\bK_{\widetilde{Z}_1} & = &
(\bK_X^*+\bK_{Z_1})^{-1}\bK_{Z_1}, \\
(\bK_X^*+\bK_{\widetilde{Z}_2})^{-1}(\bS+\bK_{\widetilde{Z}_2}) & =
& (\bK_X^*+\bK_{Z_2})^{-1}(\bS+\bK_{Z_2}). \eeqa
\end{lemma}

Combining Steps 1-5, we conclude that for any $\mu \geq 1$ and any
positive semidefinite $\bS$, a Gaussian $\bX$ is an optimal solution
of \eqref{eq:Opt-G}. This completes the direct proof of Theorem
\ref{theorem:main}.

A few comments on why we need the auxiliary optimization problem
$\widetilde{P}$ are now in place. For the classical EPI to be tight,
we need $\bK_X^*+\bK_{Z_1}$ and $\bK_X^*+\bK_{Z_2}$ to be
proportional to each other. However, by the KKT-like condition
\eqref{eq:A1}, a guarantee of proportionality requires both
multipliers $\bM_1$ and $\bM_2$ be zero. The purpose of enhancement
is to absorb the (possibly) nonzero Lagrange multipliers $\bM_1$,
$\bM_2$ into the covariance matrices of $\bZ_1$ and $\bZ_2$,
creating a new optimization problem which can be solved directly by
the classical EPI. The constant $F$ is needed to make sure that
$(P_G) = (\widetilde{P}_G)$; the choice of $F$ is motivated by the
vector Gaussian broadcast channel problem.

\section{Proof of Lemma \ref{lemma:FI}} \label{app:myFII}
We first give some preliminaries on Fisher information and score
function. This material can be found, for example, in
\cite[Chapter~3.2]{Johnson04}.

\begin{definition} \label{def:FI}
For a random vector $\bU$ with a differentiable density function
$f_U$ in $\real^n$, the Fisher information matrix $\bJ(\cdot)$ is
defined as \beq \bJ(\bU) :=
\expect[\boldsymbol{\rho}_U(\bU)\boldsymbol{\rho}_U^t(\bU)], \eeq
where the vector-valued score function $\boldsymbol{\rho}_U(\cdot)$
is defined as \beq \boldsymbol{\rho}_U(\mathbf{u}) := \nabla\log
f_U(\mathbf{u}) = \left(\frac{\partial}{\partial u_1}\log
f_U(\mathbf{u}),\cdots,\frac{\partial}{\partial u_n}\log
f_U(\mathbf{u})\right)^t. \eeq \end{definition}

The following results on score function are known.

\begin{lemma} \label{lemma:score}
The following statements on score function are true.
\begin{enumerate}
\item (Gaussian Distribution) If $\bU$ is a Gaussian vector with
zero mean and positive definite covariance matrix $\bK_U$, then \beq
\boldsymbol{\rho}_U(\mathbf{u}) = -\bK_U^{-1}\mathbf{u}.
\label{eq:C0} \eeq \item (Stein Identity) For any smooth
scalar-valued function $g$ well behaved at infinity, we have \beq
\expect[g(\bU)\boldsymbol{\rho}_U(\bU)] = -\expect[\nabla g(\bU)].
\eeq In particular, we have \beq \expect[\boldsymbol{\rho}_U(\bU)] =
0 \quad \mbox{and} \quad \expect[\bU\boldsymbol{\rho}_U^t(\bU)] =
-\bI, \eeq where $\bI$ is the identity matrix. \item (Behavior on
Convolution) If $\bU$, $\bV$ are two independent random vectors and
$\bW = \bU+\bV$, then \beq
\boldsymbol{\rho}_W(\mathbf{w})=\expect[\boldsymbol{\rho}_U(\bU)|\bW=\mathbf{w}]
=\expect[\boldsymbol{\rho}_V(\bV)|\bW=\mathbf{w}]. \eeq
\end{enumerate}
\end{lemma}

We now use the above properties of score function to prove Lemma
\ref{lemma:FI}. We first prove the Cram\'{e}r-Rao inequality. The
Fisher information matrix $\bJ(\bU)$ has nothing to do with the mean
of $\bU$, so without loss of generality we can assume that $\bU$ has
zero mean. We have \beqa \hspace{-20pt} 0 & \preceq &
\expect[(\boldsymbol{\rho}_U(\bU)+\bK_U^{-1}\bU)(\boldsymbol{\rho}_U(\bU)+\bK_U^{-1}\bU)^t] \\
\hspace{-20pt} & = &
\expect[\boldsymbol{\rho}_U(\bU)\boldsymbol{\rho}_U(\bU)^t]
+\bK_U^{-1}\expect[\bU\boldsymbol{\rho}_U^t(\bU)]
+\expect[\boldsymbol{\rho}_U(\bU)\bU^t]\bK_U^{-1} +
\bK_U^{-1}\expect[\bU\bU^t]\bK_U^{-1} \\
\hspace{-20pt} & = & \bJ(\bU)-\bK_U^{-1}-\bK_U^{-1}+\bK_U^{-1} \label{eq:CC} \\
\hspace{-20pt} & = & \bJ(\bU)-\bK_U^{-1}. \eeqa Here in
\eqref{eq:CC} we use the facts that \beq
\expect[\boldsymbol{\rho}_U(\bU)\boldsymbol{\rho}_U(\bU)^t] =
\bJ(\bU) \eeq by the definition of Fisher information matrix and
that \beq \expect[\bU\boldsymbol{\rho}_U^t(\bU)] =
\expect[\boldsymbol{\rho}_U(\bU)\bU^t] = \bI \eeq by the Stein
identity. We conclude that $\bJ(\bU) \succeq \bK_U^{-1}$ for any
random vector $\bU$ with a strictly positive definite covariance
matrix $\bK_U$.

The matrix FII can be proved similarly: \beqa \hspace{-28pt} 0 &
\preceq &
\expect[(\boldsymbol{\rho}_W(\bW)-\bA\boldsymbol{\rho}_U(\bU)-(\bI-\bA)\boldsymbol{\rho}_V(\bV))
(\boldsymbol{\rho}_W(\bW)-\bA\boldsymbol{\rho}_U(\bU)-(\bI-\bA)\boldsymbol{\rho}_V(\bV))^t]
\\ \hspace{-28pt}
& = & \expect[\boldsymbol{\rho}_W(\bW)\boldsymbol{\rho}_W^t(\bW)]
+\bA\expect[\boldsymbol{\rho}_U(\bU)\boldsymbol{\rho}_U^t(\bU)]\bA^t
+(\bI-\bA)\expect[\boldsymbol{\rho}_V(\bV)\boldsymbol{\rho}_V^t(\bV)](\bI-\bA)^t
\nonumber \\ \hspace{-28pt} & & \hspace{20pt}
-\expect[\boldsymbol{\rho}_W(\bW)\boldsymbol{\rho}_U^t(\bU)]\bA^t
-\bA\expect[\boldsymbol{\rho}_U(\bU)\boldsymbol{\rho}_W^t(\bW)]
\nonumber \\ \hspace{-28pt} & & \hspace{20pt}
-\expect[\boldsymbol{\rho}_W(\bW)\boldsymbol{\rho}_V^t(\bV)](\bI-\bA)^t
-(\bI-\bA)\expect[\boldsymbol{\rho}_V(\bV)\boldsymbol{\rho}_W^t(\bW)]
\nonumber \\ \hspace{-28pt} & & \hspace{20pt}
+\bA\expect[\boldsymbol{\rho}_U(\bU)\boldsymbol{\rho}_V^t(\bV)](\bI-\bA)^t
+(\bI-\bA)\expect[\boldsymbol{\rho}_V(\bV)\boldsymbol{\rho}_U^t(\bU)]\bA^t.
\label{eq:CC0} \eeqa By the definition of Fisher information matrix,
\beq
\expect[\boldsymbol{\rho}_W(\bW)\boldsymbol{\rho}_W^t(\bW)]=\bJ(\bW),
\quad
\expect[\boldsymbol{\rho}_U(\bU)\boldsymbol{\rho}_U^t(\bU)]=\bJ(\bU),
\quad
\expect[\boldsymbol{\rho}_V(\bV)\boldsymbol{\rho}_V^t(\bV)]=\bJ(\bV).
\label{eq:CC1} \eeq By the convolution behavior of score function,
\beq \expect[\boldsymbol{\rho}_W(\bW)\boldsymbol{\rho}_U^t(\bU)]=
\expect[\boldsymbol{\rho}_W(\bW)\expect[\boldsymbol{\rho}_U^t(\bU)|\bW]]
= \expect[\boldsymbol{\rho}_W(\bW)\boldsymbol{\rho}_W^t(\bW)] =
\bJ(\bW) \label{eq:CC12} \eeq and similarly \beq
\expect[\boldsymbol{\rho}_U(\bU)\boldsymbol{\rho}_W^t(\bW)] =
\bJ(\bW), \quad
\expect[\boldsymbol{\rho}_W(\bW)\boldsymbol{\rho}_V^t(\bV)] =
\expect[\boldsymbol{\rho}_V(\bV)\boldsymbol{\rho}_W^t(\bW)] =
\bJ(\bW). \label{eq:CC2} \eeq Finally, since $\bU$, $\bV$ are
independent and by the Stein identity with $f=1$, we have \beq
\expect[\boldsymbol{\rho}_U(\bU)\boldsymbol{\rho}_V^t(\bV)]
=\expect[\boldsymbol{\rho}_U(\bU)]\expect[\boldsymbol{\rho}_U^t(\bV)]
= 0 \eeq and similarly \beq
\expect[\boldsymbol{\rho}_V(\bV)\boldsymbol{\rho}_U^t(\bU)]=0.
\label{eq:CC3} \eeq Substituting \eqref{eq:CC1}-\eqref{eq:CC3} into
\eqref{eq:CC0}, we obtain \beqa \hspace{-20pt} 0 & \preceq &
\bJ(\bW)+\bA\bJ(\bU)\bA^t+(\bI-\bA)\bJ(\bV)(\bI-\bA)^t-\bJ(\bW)\bA^t
-\bA\bJ(\bW) -\bJ(\bW)(\bI-\bA)^t \nonumber \\ & & \hspace{20pt} -(\bI-\bA)\bJ(\bW) \\
\hspace{-20pt} & = &
-\bJ(\bW)+\bA\bJ(\bU)\bA^t+(\bI-\bA)\bJ(\bV)(\bI-\bA)^t, \eeqa which
gives \beq \bJ(\bW) \preceq
\bA\bJ(\bU)\bA^t+(\bI-\bA)\bJ(\bV)(\bI-\bA)^t \eeq for any square
matrix $\bA$. This completes the proof.

\section{A Perturbation Proof of Theorem \ref{theorem:main}}
\label{app:perturbproof} We first give the outline of the proof.

\emph{Proof Outline.} Without loss of generality, let us assume that
$\bS \succ 0$. To show that a Gaussian $\bX$ is an optimal solution
of $P$, it is sufficient to show that $(P)=(P_G)$. We have $(P) \geq
(P_G)$ (for free); we only need to show that $(P) \leq (P_G)$. For
that purpose we shall consider the auxiliary optimization problem
$\overline{P}$: \beq
\begin{array}{ll} \max_{p(\mathbf{x})} & h(\bX+\widetilde{\bZ}_1)-\mu
h(\bX+\bZ_2)+h(\bZ_1)-h(\widetilde{\bZ}_1) \\
\mbox{subject to} & \Cov(\bX) \preceq \bS, \end{array}
\label{opt:aux} \eeq where the maximization is over all random
vector $\bX$ independent of $\widetilde{\bZ}_1$ and $\bZ_2$.
Compared with the auxiliary optimization problem $\widetilde{P}$ in
the direct proof, this enhancement is only on $\bZ_1$. Following the
same footsteps as those in the direct proof, we can show that $(P)
\leq (\overline{P})$ and $(\overline{P}_G) = (P_G)$. (In proving
$(P) \leq (\overline{P})$, only the equations
\eqref{ww}-\eqref{eq:A7} and the Markov chain \eqref{ww2} in
Appendix \ref{app:basicproof} are needed.) All we need to show now
is that $ (\overline{P})=(\overline{P}_G)$.

\begin{figure}
\centering \centering \epsfxsize 3 in \epsfbox{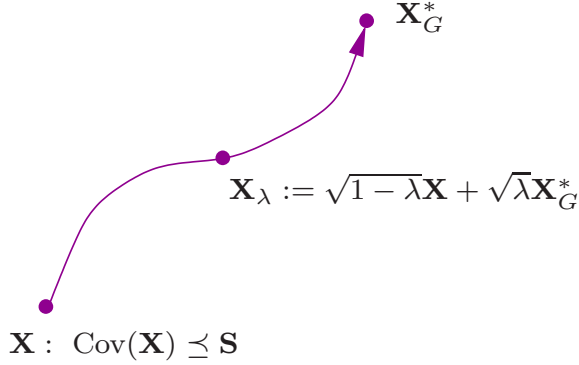} \caption{A
monotone path connecting $\bX$ and $\bX_G^*$.} \label{fig:mono}
\end{figure}

\emph{Proof of $(\overline{P})=(\overline{P}_G)$.} To show that
$(\overline{P})=(\overline{P}_G)$, we shall show that $\bX_G^*$ is a
global optimal solution of $\overline{P}$. For that we shall prove
the following strong result: for any admissible random vector $\bX$
there is a monotone increasing path connecting $\bX$ and $\bX_G^*$
(see Figure \ref{fig:mono}).

We consider the ``covariance-preserving" transformation of Dembo et
al. \cite{DCT91}: \beq \bX_\lambda =
\sqrt{1-\lambda}\bX+\sqrt{\lambda}\bX_G^*, \quad \lambda \in [0,1].
\eeq Then $\{\bX_{\lambda}\}$ is a family of distributions indexed
by $\lambda \in [0,1]$ and connecting $\bX$ (when $\lambda=0$) with
$\bX_G^*$ (when $\lambda=1$). Let $\overline{g}(\lambda)$ be the
objective function of $\overline{P}$ evaluated along the path
$\{\bX_\lambda\}$: \beq \overline{g}(\lambda) :=
h(\bX_\lambda+\widetilde{\bZ}_1)-\mu
h(\bX_\lambda+\bZ_2)+h(\bZ_1)-h(\widetilde{\bZ}_1). \eeq Next, we
calculate the derivative of $\overline{g}$ over $\lambda$.

Note that $\widetilde{\bZ}_1$ is Gaussian and that a Gaussian
distribution is stable under convolution. We can write \beq
\widetilde{\bZ}_1 =
\sqrt{1-\lambda}\widetilde{\bZ}_{1,1}+\sqrt{\lambda}\widetilde{\bZ}_{1,2},
\eeq where $\widetilde{\bZ}_{1,1}$, $\widetilde{\bZ}_{1,2}$ are
independent and have the same distribution as that of
$\widetilde{\bZ}_1$. We have \beqa h(\bX_\lambda+\widetilde{\bZ}_1)
& = & h(\sqrt{1-\lambda}\bX+\sqrt{\lambda}\bX_G^*+\widetilde{\bZ}_i)
\\ & = & h(\sqrt{1-\lambda}(\bX+\widetilde{\bZ}_{1,1})+\sqrt{\lambda}(\bX_G^*
+\widetilde{\bZ}_{1,2})) \\ & = &
h(\bX+\widetilde{\bZ}_{1,1}+\sqrt{\lambda(1-\lambda)^{-1}}(\bX_G^*
+\widetilde{\bZ}_{1,2})) +(n/2)\log(1-\lambda). \eeqa By the
(vector) de Bruijn identity \cite[Theorem~14]{DCT91}, \beqa
\hspace{-40pt} &
& 2(1-\lambda)\frac{d}{d\lambda}h(\bX_\lambda+\widetilde{\bZ}_1) \nonumber \\
\hspace{-40pt} & & \hspace{20pt} = \;
(1-\lambda)^{-1}\Tr\left((\bK_X^*+\bK_{\widetilde{Z}_1})\bJ\left(\bX+\widetilde{\bZ}_{1,1}+
\sqrt{\lambda(1-\lambda)^{-1}}(\bX_G^*+\widetilde{\bZ}_{1,2})\right)\right)-n
\\ \hspace{-40pt} & & \hspace{20pt} = \;
\Tr\left((\bK_X^*+\bK_{\widetilde{Z}_1})\bJ\left(\sqrt{1-\lambda}(\bX+\widetilde{\bZ}_{1,1})+
\sqrt{\lambda}(\bX_G^*+\widetilde{\bZ}_{1,2})\right)\right)-n
\\ \hspace{-40pt} & & \hspace{20pt} = \;
\Tr\left((\bK_X^*+\bK_{\widetilde{Z}_1})\bJ\left(\bX_\lambda+\widetilde{\bZ}_1\right)\right)-n.
\label{eq:DD0} \eeqa Similarly, we have \beq
2(1-\lambda)\frac{d}{d\lambda}h(\bX_\lambda+{\bZ}_2) =
\Tr\left((\bK_X^*+\bK_{{Z}_2})\bJ\left(\bX_\lambda+{\bZ}_2\right)\right)-n.
\label{eq:DD1} \eeq Combining \eqref{eq:DD0} and \eqref{eq:DD1}, we
have \beq 2(1-\lambda)\overline{g}'(\lambda) =
\Tr\left((\bK_X^*+\bK_{\widetilde{Z}_1})\bJ(\bX_\lambda+\widetilde{\bZ}_1)
-\mu(\bK_X^*+\bK_{Z_2})\bJ(\bX_\lambda+\bZ_2)\right)+n(\mu-1).
\label{eq:D0} \eeq

By the definition of $\bK_{\widetilde{Z}_1}$ and the KKT-like
condition \eqref{eq:A1}, we have \beq
\frac{1}{2}(\bK_X^*+\bK_{\widetilde{Z}_1})^{-1} = \frac{\mu}{2}
(\bK_X^*+\bK_{Z_2})^{-1}+\bM_2. \label{eq:M} \eeq By the facts that
$\mu \geq 1$ and $\bM_2 \succeq 0$, we obtain from \eqref{eq:M} that
\beq \frac{1}{2}(\bK_X^*+\bK_{\widetilde{Z}_1})^{-1} \succeq
\frac{1}{2}(\bK_X^*+\bK_{Z_2})^{-1} \label{eq:M2} \eeq and hence
that \beq \bK_{Z_2} \succeq \bK_{\widetilde{Z}_1}. \eeq We can now
write $\bZ_2 = \widetilde{\bZ}_1 + \overline{\bZ}$, where
$\overline{\bZ}$ is Gaussian and independent of $\widetilde{\bZ}_1$.
Applying the matrix FII of Lemma \ref{lemma:FI} with \beq \bA =
(\bK_X^*+\bK_{Z_2})^{-1}(\bK_X^*+\bK_{\widetilde{Z}_1}) \quad
\mbox{and} \quad \bI-\bA =
(\bK_X^*+\bK_{Z_2})^{-1}\bK_{\overline{Z}}, \eeq we have \beqa
\hspace{-30pt}
\bJ(\bX_\lambda+\bZ_2) & = & \bJ(\bX_\lambda+\widetilde{\bZ}_1+\overline{\bZ}) \\
& \leq &
(\bK_X^*+\bK_{Z_2})^{-1}(\bK_X^*+\bK_{\widetilde{Z}_1})\bJ(\bX_\lambda+\widetilde{\bZ}_1)
(\bK_X^*+\bK_{\widetilde{Z}_1})(\bK_X^*+\bK_{Z_2})^{-1} \nonumber
\\ & & \hspace{20pt}+\;(\bK_X^*+\bK_{Z_2})^{-1}\bK_{\overline{Z}}\bJ(\overline{\bZ})
\bK_{\overline{Z}}(\bK_X^*+\bK_{Z_2})^{-1} \\ & = &
(\bK_X^*+\bK_{Z_2})^{-1}(\bK_X^*+\bK_{\widetilde{Z}_1})\bJ(\bX_\lambda+\widetilde{\bZ}_1)
(\bK_X^*+\bK_{\widetilde{Z}_1})(\bK_X^*+\bK_{Z_2})^{-1} \nonumber
\\ & & \hspace{20pt}+\;
(\bK_X^*+\bK_{Z_2})^{-1}\bK_{\overline{Z}}(\bK_X^*+\bK_{Z_2})^{-1},
\label{eq:D100} \eeqa where the last equality follows from the fact
that $\overline{\bZ}$ is Gaussian so \beq
\bK_{\overline{Z}}\bJ(\overline{\bZ}) = \bI. \eeq Substituting
\eqref{eq:D100} into \eqref{eq:D0} and using the fact that
$\bK_{\overline{Z}} = \bK_{Z_2}-\bK_{\widetilde{Z}_1}$, we obtain
\beqa
\hspace{-30pt} & & 2(1-\lambda)\overline{g}'(\lambda) \nonumber \\
\hspace{-30pt} & & \hspace{20pt} \geq
\;\Tr((\bK_X^*+\bK_{\widetilde{Z}_1})\bJ(\bX_\lambda+\widetilde{\bZ}_1)
-\mu(\bK_X^*+\bK_{\widetilde{Z}_1})\bJ(\bX_\lambda+\widetilde{\bZ}_1)
(\bK_X^*+\bK_{\widetilde{Z}_1})(\bK_X^*+\bK_{Z_2})^{-1} \nonumber
\\ \hspace{-30pt} & & \hspace{60pt}
-\,\mu\bK_{\overline{Z}}(\bK_X^*+\bK_{Z_2})^{-1})+n(\mu-1)
\\ \hspace{-30pt} & & \hspace{20pt} = \; 2\,\Tr\left(
\left((\bK_X^*+\bK_{\widetilde{Z}_1})\bJ(\bX_\lambda+\widetilde{\bZ}_1)(\bK_X^*+\bK_{\widetilde{Z}_1})
-(\bK_X^*+\bK_{\widetilde{Z}_1})\right)\bM_2\right), \label{D200}
\eeqa where the equality follows from \eqref{eq:M}. Further by the
Cram\'{e}r-Rao inequality of Lemma \ref{lemma:FI}, \beqa
\hspace{-40pt} & &
(\bK_X^*+\bK_{\widetilde{Z}_1})\bJ(\bX_\lambda+\widetilde{\bZ}_1)(\bK_X^*+\bK_{\widetilde{Z}_1})
-(\bK_X^*+\bK_{\widetilde{Z}_1}) \nonumber \\ \hspace{-40pt} & &
\hspace{20pt} \succeq \; (\bK_X^*+\bK_{\widetilde{Z}_1})
\Cov^{-1}(\bX_\lambda+\widetilde{\bZ}_1)(\bK_X^*+\bK_{\widetilde{Z}_1})
-(\bK_X^*+\bK_{\widetilde{Z}_1}) \\ \hspace{-40pt} & & \hspace{20pt}
= \;
(\bK_X^*+\bK_{\widetilde{Z}_1})\left((1-\lambda)\Cov(\bX)+\lambda\bK_X^*+\bK_{\widetilde{Z}_1}\right)^{-1}
(\bK_X^*+\bK_{\widetilde{Z}_1}) -(\bK_X^*+\bK_{\widetilde{Z}_1}) \\
\hspace{-40pt} & & \hspace{20pt} \succeq
(\bK_X^*+\bK_{\widetilde{Z}_1})\left((1-\lambda)\bS+\lambda\bK_X^*+\bK_{\widetilde{Z}_1}\right)^{-1}
(\bK_X^*+\bK_{\widetilde{Z}_1}) -(\bK_X^*+\bK_{\widetilde{Z}_1}) \\
\hspace{-40pt} & & \hspace{20pt} = -(1-\lambda)
(\bK_X^*+\bK_{\widetilde{Z}_1})\left((1-\lambda)\bS+\lambda\bK_X^*+\bK_{\widetilde{Z}_1}\right)^{-1}
(\bS-\bK_X^*). \label{D300} \eeqa Substitute \eqref{D300} into
\eqref{D200} and recall from the KKT-like condition \eqref{eq:A3}
that $(\bS-\bK_X^*)\bM_2=0$. We have \beqa \hspace{-40pt} & &
\Tr\left(\left((\bK_X^*+\bK_{\widetilde{Z}_1})\bJ(\bX_\lambda+\widetilde{\bZ}_1)
(\bK_X^*+\bK_{\widetilde{Z}_1})
-(\bK_X^*+\bK_{\widetilde{Z}_1})\right)\bM_2\right) \nonumber \\
\hspace{-40pt} & & \hspace{20pt} \geq \; -(1-\lambda)\,\Tr\left(
(\bK_X^*+\bK_{\widetilde{Z}_1})\left((1-\lambda)\bS+\lambda\bK_X^*+\bK_{\widetilde{Z}_1}\right)^{-1}
(\bS-\bK_X^*)\bM_2\right) \\ \hspace{-40pt} & & \hspace{20pt} = \;
0. \eeqa We conclude that \beq \overline{g}'(\lambda) \geq 0, \quad
\forall \, \lambda \in [0,1], \eeq i.e., $\{\bX_\lambda\}$ is a
monotone increasing path connecting $\bX$ and $\bX_G^*$. We have
found a monotone increasing path for every admissible $\bX$, so
$\bX_G^*$ is an optimal solution of $\overline{P}$. This completes
the perturbation proof of Theorem \ref{theorem:main}.

A few comments on the difference between the direct proof and the
perturbation proof of Theorem \ref{theorem:main} are now in place.
In the direct proof, we enhance both $\bZ_1$ and $\bZ_2$ to obtain
the proportionality so that the classical EPI can be applied to
solve the auxiliary optimization problem $\widetilde{P}$. For the
perturbation proof, however, we only need to enhance $\bZ_1$. (If
the lower constraint $\bK_X \succeq 0$ does not bite, i.e.,
$\bM_1=0$, no enhancement is needed at all.) A direct perturbation
is then used to show that $\bX_G^*$ is an optimal solution of the
auxiliary optimization problem $\overline{P}$. Neither the classical
EPI nor the worst noise result of Lemma \ref{lemma:worst} is needed
in the perturbation proof.

\section{Proof of Corollary \ref{cor:es}} \label{app:es}
For any random vector $\bX$ in $\real^2$ such that $\Cov(\bX) \leq
\bS$ and any $\mu \geq 1$, we have from Theorem \ref{theorem:main}
that \beq h(\bX+\bZ_1)-\mu h(\bX+\bZ_2) \leq
\max_{\mathbf{0}\,\preceq \,\bK_X \preceq \,\bS}\left\{
\frac{1}{2}\log((2\pi e)^2|\bK_X+\bK_{Z_1}|)-\frac{\mu}{2}\log((2\pi
e)^2|\bK_X+\bK_{Z_2}|)\right\}. \label{eq:E1} \eeq Adding a constant
term $\mu h(\bZ_2)-h(\bZ_1)$ to both sides of \eqref{eq:E1}, we
obtain \beq I(\bX;\bX+\bZ_1)-\mu I(\bX;\bX+\bZ_2) \leq
\max_{\mathbf{0}\,\preceq \,\bK_X \preceq \,\bS}\left\{
\frac{1}{2}\log\left|\bI+\bK_{Z_1}^{-1}\bK_X\right|
-\frac{\mu}{2}\log\left|\bI+\bK_{Z_2}^{-1}\bK_X\right|\right\}.
\label{eq:E2} \eeq Let
$\bK_{Z_i}=\bV_i\boldsymbol{\Sigma}_i\bV_i^t$, where
$\bV_i=(\bv_{i1},\bv_{i2})$ is an orthonormal matrix and
$\boldsymbol{\Sigma}_i=\Diag(\lambda_{i1},\lambda_{i2})$ is a
diagonal matrix. Next, we consider taking the limits of both sides
of \eqref{eq:E2} as $\lambda_{12},\lambda_{21} \rightarrow \infty$.

First consider the limit of the left-hand side of \eqref{eq:E2}. We
need the following simple lemma.

\begin{lemma} \label{lemma:es} Let $\bZ=(Z_1,Z_2)^t$ where
$Z_1$, $Z_2$ are two independent Gaussian variables with variance
$\sigma_1^2$ and $\sigma_2^2$, respectively. For any random vector
$\bX=(X_1,X_2)^t$ with finite variances and independent of $\bZ$, we
have \beq \lim_{\sigma_2^2 \rightarrow
\infty}I(\bX;\bX+\bZ)=I(X_1;X_1+Z_1). \label{eq:es} \eeq
\end{lemma}

\emph{Proof.} By the chain rule of mutual information, \beq
I(\bX;\bX+\bZ) =
I(X_1;X_1+Z_1)+I(X_1;X_2+Z_2|X_1+Z_1)+I(X_2;\bX+\bZ|X_1). \label{EE}
\eeq Due to the Markov chains $X_1+Z_1 \rightarrow X_1 \rightarrow
X_2+Z_2$ and $X_1 \rightarrow X_2 \rightarrow X_2+Z_2$, we have \beq
I(X_1;X_2+Z_2|X_1+Z_1) \leq I(X_1;X_2+Z_2) \leq I(X_2;X_2+Z_2).
\label{eq:E2.5} \eeq Furthermore, we have \beqa I(X_2;\bX+\bZ|X_1) &
= & I(X_2;X_2+Z_2|X_1)+I(X_2;X_1+Z_1|X_1,X_2+Z_2) \\
& = & I(X_2;X_2+Z_2|X_1)+I(X_2;Z_1|X_1,X_2+Z_2) \\
& = & I(X_2;X_2+Z_2|X_1) \label{eq:E3} \\ & \leq & I(X_2;X_2+Z_2),
\label{eq:E4} \eeqa where \eqref{eq:E3} follows from the fact that
$Z_1$ is independent of $Z_2$ and $\bX$ so
$I(X_2;Z_1|X_1,X_2+Z_2)=0$, and \eqref{eq:E4} is due to the Markov
chain $X_1 \rightarrow X_2 \rightarrow X_2+Z_2$. Note that \beq
\lim_{\sigma_2^2 \rightarrow \infty}I(X_2;X_2+Z_2) \leq
\lim_{\sigma_2^2 \rightarrow \infty}
\frac{1}{2}\log\left(1+\frac{\Var(X_2)}{\sigma_2^2}\right)=0 \eeq
with finite $\Var(X_2)$. We thus have from \eqref{eq:E2.5} and
\eqref{eq:E4} that both $I(X_1;X_2+Z_2|X_1+Z_1)$ and
$I(X_2;\bX+\bZ|X_1)$ tend to zero in the limit as $\sigma_2^2
\rightarrow \infty$. The desired result \eqref{eq:es} follows by
taking the limit $\sigma_2^2 \rightarrow \infty$ on both sides of
\eqref{EE}, which completes the proof. \hfill $\square$

Let
$\overline{\bZ}_i=(\overline{Z}_{i,1},\overline{Z}_{i,2})^t=\bV_i^t\bZ_i$.
Then, $\overline{Z}_{i,1}$ and $\overline{Z}_{i,2}$ are independent.
By Lemma \ref{lemma:es}, \beqa \lim_{\lambda_{12} \rightarrow
\infty}I(\bX;\bX+\bZ_1) = \lim_{\lambda_{12} \rightarrow
\infty}I(\bV_1^t\bX;\bV_1^t\bX+\bV_1^t\bZ_1) =
I(\bv_{11}^t\bX;\bv_{11}^t\bX+\overline{Z}_{11}) \\
\lim_{\lambda_{21} \rightarrow \infty}I(\bX;\bX+\bZ_2) =
\lim_{\lambda_{21} \rightarrow
\infty}I(\bV_2^t\bX;\bV_2^t\bX+\bV_2^t\bZ_1) =
I(\bv_{22}^t\bX;\bv_{22}^t\bX+\overline{Z}_{22}), \eeqa which gives
\beq \lim_{\lambda_{12},\lambda_{21} \rightarrow
\infty}I(\bX;\bX+\bZ_1)-\mu
I(\bX+\bZ_2)=I(\bv_{11}^t\bX;\bv_{11}^t\bX+\overline{Z}_{11})- \mu
I(\bv_{22}^t\bX;\bv_{22}^t\bX+\overline{Z}_{22}). \label{eq:E5} \eeq

Next, we consider the limit of the right-hand side of \eqref{eq:E2}.
For any semidefinite $\bK_X$, we have \beqa & &
\lim_{\lambda_{12},\lambda_{21} \rightarrow \infty}
\left\{\frac{1}{2}\log\left|\bI+\bK_{Z_1}^{-1}\bK_X\right|
-\frac{\mu}{2}\log\left|\bI+\bK_{Z_2}^{-1}\bK_X\right|\right\}
\nonumber \\
& & \hspace{20pt} =\;\lim_{\lambda_{12},\lambda_{21} \rightarrow
\infty}\left\{
\frac{1}{2}\log\left|\bI+\boldsymbol{\Sigma}_1^{-1}\bV_1^t\bK_X\bV_1\right|
-\frac{\mu}{2}\log\left|\bI+\boldsymbol{\Sigma}_2^{-1}\bV_2^t\bK_X\bV_2\right|\right\}
\\
& & \hspace{20pt}
=\;\frac{1}{2}\log\left(1+\lambda_{11}^{-1}\bv_{11}^t\bK_X\bv_{11}\right)-
\frac{\mu}{2}\log\left(1+\lambda_{22}^{-1}\bv_{22}^t\bK_X\bv_{22}\right)
\label{EE2} \eeqa due to the continuity of $\log|\bI+\bA|$ over the
semidefinite $\bA$. Moreover, the convergence of \eqref{EE2} is
\emph{uniform} in $\bK_X$, because the continuity of $\log|\bI+\bA|$
over $\bA$ is uniform and $\bV_i^t\bK_X\bV_i$, $i=1,2$, are bounded
for $0 \preceq \bK_X \preceq \bS$. we thus have \beqa & &
\lim_{\lambda_{12},\lambda_{21} \rightarrow \infty}\left\{
\max_{\mathbf{0}\,\preceq \,\bK_X \preceq \, \bS}\left\{
\frac{1}{2}\log\left|\bI+\bK_{Z_1}^{-1}\bK_X\right|
-\frac{\mu}{2}\log\left|\bI+\bK_{Z_2}^{-1}\bK_X\right|\right\}\right\}
\nonumber \\
& & \hspace{20pt} =\;\max_{\mathbf{0}\,\preceq \,\bK_X \preceq
\,\bS}\left\{
\frac{1}{2}\log\left(1+\lambda_{11}^{-1}\bv_{11}^t\bK_X\bv_{11}\right)-
\frac{\mu}{2}\log\left(1+\lambda_{22}^{-1}\bv_{22}^t\bK_X\bv_{22}\right)\right\}.
\label{eq:E6} \eeqa

Substituting \eqref{eq:E5} and \eqref{eq:E6} into \eqref{eq:E1},
we obtain \beqa & &
I(\bv_{11}^t\bX;\bv_{11}^t\bX+\overline{Z}_{11})-\mu
I(\bv_{22}^t\bX;\bv_{22}^t\bX+\overline{Z}_{22}) \nonumber \\
&& \hspace{20pt} \leq \; \max_{\mathbf{0}\,\preceq \,\bK_X \preceq
\,\bS}\left\{
\frac{1}{2}\log\left(1+\lambda_{11}^{-1}\bv_{11}^t\bK_X\bv_{11}\right)-
\frac{\mu}{2}\log\left(1+\lambda_{22}^{-1}\bv_{22}^t\bK_X\bv_{22}\right)\right\}
\eeqa and hence \beqa \hspace{-40pt} & &
h(\bv_{11}^t\bX+\overline{Z}_{11})-\mu
h(\bv_{22}^t\bX+\overline{Z}_{22}) \nonumber \\
\hspace{-40pt} && \hspace{20pt} \leq \; \max_{\mathbf{0}\,\preceq
\,\bK_X \preceq \,\bS} \left\{\frac{1}{2}\log\left(2\pi
e\left(\bv_{11}^t\bK_X\bv_{11}+\lambda_{11}\right)\right)-
\frac{\mu}{2}\log\left(2\pi e
\left(\bv_{22}^t\bK_X\bv_{22}+\lambda_{22}\right)\right)\right\}
\label{eq:E7} \eeqa for any random vector $\bX$ such that $\Cov(\bX)
\preceq \bS$ and any $\mu \geq 1$. This completes the proof.

\section{Proof of Corollary \ref{cor:LV}} \label{app:LV}
Let $\bv_1=(1,1)^t$ and $\bv_2=(0,1)^t$. Consider $
\left\{\bX:\,\Var(X_1) \leq a_1\right\} =
\bigcup_{\bS}\left\{\bX:\,\Cov(\bX) \preceq \bS\right\}$ where the
union is over all $\bS$ such that $(\bS)_{11}=a_1$. By Corollary
\ref{cor:es}, a Gaussian $(X_1,X_2$ is an optimal solution to the
optimization problem \beq
\begin{array}{ll}
\max_{p(x_1,x_2)} & h(X_1+X_2+Z)-\mu h(X_2+Z) \\
\mathrm{subject\;to} & \Var(X_1) \leq a_1,
\end{array} \label{opt:LV2} \eeq where $\mu \geq 1$, and the maximization is over all
jointly distributed random variables $(X_1,X_2)$ independent of $Z$.
Let $(X_{1G}^*,X_{2G}^*)$ be the Gaussian optimal solution of the
\eqref{opt:LV2}. Then \beq h(X_1+X_2+Z)-\mu h(X_2+Z) \leq
h(X_{1G}^*+X_{2G}^*+Z)-\mu h(X_{2G}^*+Z) \label{eq:F1} \eeq for any
jointly distributed random variables $(X_1,X_2)$ such that
$\Var(X_1) \leq a_1$.

It is easy to verify that $h(X_{2G}^*+Z)$ is a continuous function
of $\mu$. When $\mu = \infty$, $h(X_{2G}^*+Z)=h(Z)$; when $\mu = 1$,
$h(X_{2G}^*+Z)=a_2^*$ where $a_2^*$ was defined in \eqref{eq:a2}. By
the intermediate value theorem, for any $h(Z) \leq a_2 \leq a_2^*$
there is a $\mu$ for which $h(X_{2G}^*+Z)=a_2$. Hence for any
jointly distributed random variables $(X_1,X_2)$ such that
$\Var(X_1) \leq a_1$ and $h(X_2+Z) \leq a_2$, we have by
\eqref{eq:F1} that \beqa h(X_1+X_2+Z) & \leq &
h(X_{1G}^*+X_{2G}^*+Z)+\mu\left(h(X_2+Z)-h(X_{2G}^*+Z)\right) \\
& \leq & h(X_{1G}^*+X_{2G}^*+Z). \eeqa We conclude that a Gaussian
solution is an optimal solution of \eqref{app:LV2} for any $a_1 \geq
0$ and any $h(Z) \leq a_2 \leq a_2^*$. This completes the proof.

\section{Proof of the Outer Bound \eqref{eq:SW}} \label{app:CEO}
Let $W_1$ and $W_2$ be the encoded messages for $\{\bY_1[m]\}$ and
$\{\bY_2[m]\}$, respectively. Let
$\underline{\bY_i^m}:=(\bY_i[1],\cdots,\bY_i[m])$ and
$U[m]:=(W_2,\underline{\bY_1^{m-1}})$. We have
\beqa NR_2 & = & H(W_2) \\
& \geq & H(W_2)-H(W_2|\underline{\bY_2^N}) \\
& = & I(W_2;\underline{\bY_2^N}) \\
& = & \sum_{m=1}^N I(W_2;\bY_2[m]|\underline{\bY_2^{m-1}}) \\
& = & \sum_{m=1}^N
\left(h(\bY_2[m]|\underline{\bY_2^{m-1}})-h(\bY_2[m]|W_2,\underline{\bY_2^{m-1}})\right)
\\
& = & \sum_{m=1}^N
\left(h(\bY_2[m])-h(\bY_2[m]|W_2,\underline{\bY_2^{m-1}},\underline{\bY_1^{m-1}})\right)
\label{eq:deg} \\
& \geq & \sum_{m=1}^N
\left(h(\bY_2[m])-h(\bY_2[m]|W_2,\underline{\bY_1^{m-1}})\right) \\
& = & \sum_{m=1}^N I(U[m];\bY_2[m]), \label{F1} \eeqa where
\eqref{eq:deg} follows from the fact that $\underline{\bY_1^{m-1}}$
is a degraded version of $\underline{\bY_2^{m-1}}$ for
$m=1,\cdots,N$. Furthermore,
\beqa NR_1 & = & H(W_1) \\
& \geq & H(W_1)-H(W_1|W_2,\underline{\bY_1^N}) \\
& = & I(W_1;W_2,\underline{\bY_1^N}) \\
& = & I(W_1;\underline{\bY_1^N}|W_2) + I(W_1;W_2) \\
& \geq & I(W_1;\underline{\bY_1^N}|W_2) \\
& = & \sum_{m=1}^N I(W_1;\bY_1[m]|W_2,\underline{\bY_1^{m-1}})
\\ & = & \sum_{m=1}^N I(W_1,\widehat{\bY}_1[m];\bY_1[m]|W_2,\underline{\bY_1^{m-1}})
\label{eq:deg2} \\ & \geq & \sum_{m=1}^N
I(\widehat{\bY}_1[m];\bY_1[m]|W_2,\underline{\bY_1^{m-1}})
\\ & = & \sum_{m=1}^N I(\widehat{\bY}_1[m];\bY_1[m]|U[m]) \label{F2} \eeqa where
\eqref{eq:deg2} follows from the Markov chain $\bY_1[m] \rightarrow
(W_1,W_2) \rightarrow \widehat{\bY}_1[m]$ for $m=1,\cdots,N$.
Finally, let $Q$ be a random variable uniformly distributed over
$\{1,\cdots,N\}$ and independent of any other random
variables/vectors. We have from \eqref{F1} and \eqref{F2} that \beq
R_1 \geq I(U[Q];\bY_2[Q]|Q) = I(U[Q],Q;\bY_2[Q])-I(Q;\bY_2[Q]) =
I(U[Q],Q;\bY_2[Q])=I(U;\bY_2) \eeq and that \beq R_2 \geq
I(\widehat{\bY}_1[Q];\bY_1[Q]|U[Q],Q)=I(\widehat{\bY}_1;\bY_1|U)
\eeq by defining \beq U:=(Q,U[Q]),\quad
\widehat{\bY}_1:=\widehat{\bY}_1[Q],\quad \bY_1:=\bY_1[Q],\quad
\bY_2:=\bY_2[Q]. \eeq For each $m=1,\cdots,N$, \beq
\bY_1[m]=\bY_2[m]+\bZ[m] \rightarrow \bY_2[m] \rightarrow
U[m]=(W_2,\underline{\bY_1^{m-1}}) \eeq forms a Markov chain because
$\bZ[m]$ is independent of $(W_2,\bY_1^{m-1})$. Therefore, \beq
\bY_1 \rightarrow \bY_2 \rightarrow U \eeq also forms a Markov
chain. This completes the proof.
\end{appendix}

\section*{Acknowledgment}
The authors wish to thank both the reviewers and the Associate
Editor for their careful review of the manuscript, which has helped
to improve the technical quality of the paper.

\end{document}